\documentclass[%
 reprint,
nofootinbib,
 twocolumn,
 amsmath,amssymb,
 aps,
 usenames,dvipsnames,
pra,
longbibliography,
]{revtex4-1}

\usepackage{importsAQC}
\usepackage{booktabs}
\usepackage{appendix}

\begin{document}

\title{Approximations in transmon simulation}% Force line breaks with \\
\author{Tyler Jones$^{1,2,3,*}$}
\blfootnote{$^*$ tyler.jones@uq.edu.au}
\author{Kaiah Steven$^1$}
\author{Xavier Poncini$^1$}
\author{Matthew Rose$^1$}
\author{Arkady Fedorov$^{2,3}$ }%
\affiliation{%
 $^1$ Max Kelsen, Spring Hill, Queensland 4000, Australia\\
 $^2$ ARC Centre of Excellence for Engineered Quantum Systems, St Lucia, Queensland 4072, Australia\\
 $^3$ School of Mathematics and Physics, University of Queensland, St Lucia, Queensland 4072, Australia
}%

\date{\today}% It is always \today, today,

\begin{abstract}

Classical simulations of time-dependent quantum systems are widely used in quantum control research. In particular, these simulations are commonly used to host iterative optimal control algorithms. This is convenient for algorithms which are too onerous to run in the loop with current-day quantum hardware, as well as for researchers without consistent access to hardware. However, if the model used to represent the system is not selected carefully, an optimised control protocol may be rendered futile when applied to hardware. We present a series of models, ordered in a hierarchy of progressive approximation, which appear in quantum control literature. The validity of each model is characterised experimentally by designing and benchmarking control protocols for an \textsf{IBMQ} cloud quantum device. This result demonstrates error amplification induced by the application of a first-order perturbative approximation. Furthermore, the emergence of errors which cannot be corrected by simple amplitude scaling of control pulses is demonstrated in simulation, due to an underlying mistreatment of non-computational dynamics. Finally, an evaluation of simulated control dynamics reveals that despite the substantial variance in numerical predictions across the proposed models, the complexity of discovering local optimal control protocols appears invariant in the simple control scheme setting.

\end{abstract}

\pacs{Valid PACS appear here}% PACS, the Physics and Astronomy
                             
\maketitle

\section{\label{sec:Introduction}Introduction}

Since the turn of the century, superconducting qubits have emerged as a leading candidate for the realisation of a universal fault-tolerant quantum computer. To harness the potential of these systems, there remains an important focus on improving the speed and accuracy of superconducting qubit gate operations. This focus is particularly illustrated by recent work in the field of quantum optimal control \cite{Allen_2017,Spiteri_2018,Werninghaus_2021,Kirchhoff_2018,Wu_2020,Garcia_2020,riazOptimalControlMethods2019}, 
an area concerned with determining control protocols that minimise gate error.  Notably, state-of-the-art developments in machine learning have been adapted to the quantum control domain \cite{An_2019,Zhang2019,Xu_2019,Wauters_2020,Zahedinejad_2016,Niu_2019,Bukov_2018, Wu_2019, Abdelhafez_2020}, leading to the discovery of control schemes that result in order-of-magnitude reductions in gate infidelity~\cite{Niu_2019}. These techniques are largely deployed in classical simulations rather than on quantum hardware due to experimental constraints, including iteration speed and difficulty in extracting the quantum state. Whilst the move to simulation-based optimisation does carry significant benefits in accessibility and iteration time, the models used have to rely on abstracted representations of the relevant quantum system derived through a series of approximations. There should be due caution taken when reporting optimal control results that simplify system dynamics in such a way.

In this paper, we consider the system of a transmon qubit \cite{Koch_2007} coupled to a readout resonator in the dispersive regime. The simplicity of this design, along with its charge-noise insensitive nature, has led to its prevalence in research laboratories. The circuit architecture of the transmon is nominally identical to the Cooper pair box architecture, with the additional requirement of operation in a regime of high $E_J/E_C$ (Josephson energy and charging energy respectively). While this introduces charge noise insensitivity, it also results in a weakened anharmonicity, reducing isolation of the computational subspace from higher energy levels. This enhanced risk of leakage intrinsically reduces gate operation speed, an undesirable consequence considering the limited coherence times available in modern devices. The dichotomy of suppressing leakage error and accelerating operation speed has led to the discovery of several control schemes designed to exploit nuances in system dynamics, most notably the derivative removal by adiabatic gate (DRAG) pulse \cite{motzoi_gambetta_rebentrost_wilhelm_2009}. 

The problem of finding optimal quantum control schemes is often addressed algorithmically, whether through standard gradient-based approaches like Krotov, GRAPE or GOAT \cite{Reich_2012, Machnes_2018, Khaneja_2004}, or through the application of neural network-based algorithms \cite{Niu_2019, Bukov_2018, Wu_2019}. A combination of the derivative computations and magnitude of iterations required for these algorithmic approaches frequently limits them to operation within simulations, rather than closed-loop optimisation. Given this, it is imperative that relevant dynamics of the quantum system are not erased through model abstraction.

\begin{figure}
    \centering
    \includegraphics[width=0.93\linewidth]{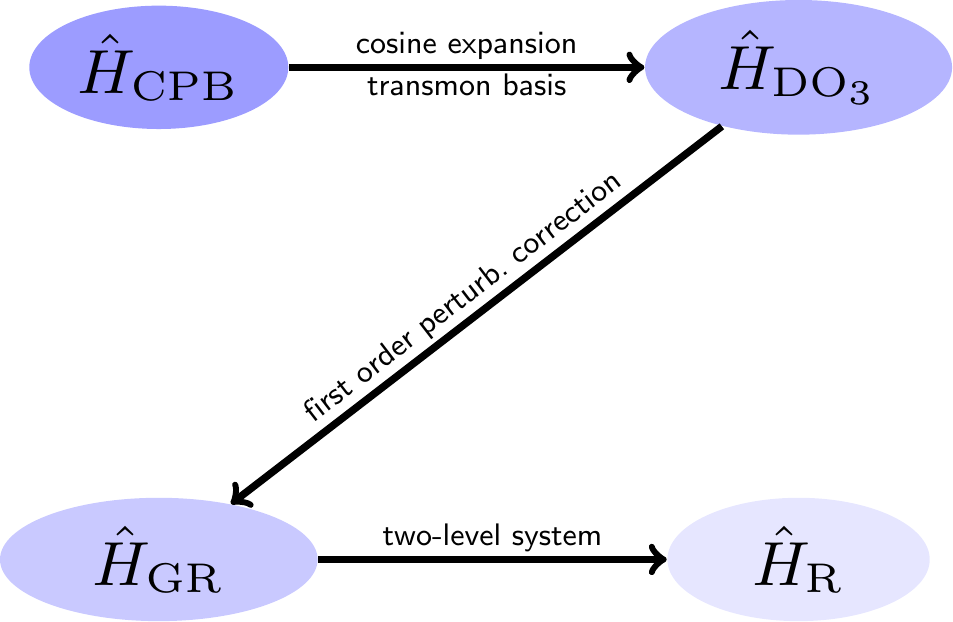}
    % \scalebox{0.85}{\HamiltonianHiarchyIInew}
    \caption{A hierarchy of system Hamiltonians, ordered by level of approximation, is proposed. $H_{\mathrm{CPB}}$ represents the model which most accurately describes a coupled transmon-resonator system. Following an arrow indicates the application of an approximation which obfuscates the system dynamics in some manner.}
    \label{fig:hierarchy}
\end{figure}

This research seeks to examine the validity of a series of approximations made to transmon-resonator systems, especially in the context of single qubit control. Simulation environments will be developed for a defined hierarchy of Hamiltonian models as seen in Fig. \ref{fig:hierarchy}. These environments will be configurable to match the energy parameters of any coupled transmon-resonator system, permitting the simulation of \textsf{IBMQ}'s  Qiskit Pulse compatible backends. This will allow for the contrast of system properties at each level of approximation, exposing the ramifications of each simplification. A set of control parameters will be optimised to design a given gate using each environment, and validated through deployment to the \textsf{IBMQ} device \texttt{ibmq\_armonk} to illustrate the enhanced fidelity induced by using the more faithful model. Noting that the particular error type demonstrated on the \textsf{IBMQ} device can be mitigated by parameter calibration in experiment, model disparities in the frequency domain are presented in order to highlight a more pathological set of deviations that are not able to be resolved through simple experimental calibration. Finally, simulated control landscapes of each model will be contrasted and traversed in order to quantify the ease of finding optimal controls for each system. It is anticipated that these results will emphasise the importance optimal control researchers should place on faithful system modelling.

\section{\label{sec:Theory} Transmon Theory}
\begin{figure}[!hbpt]
    \centering
    \includegraphics[width=0.73\linewidth]{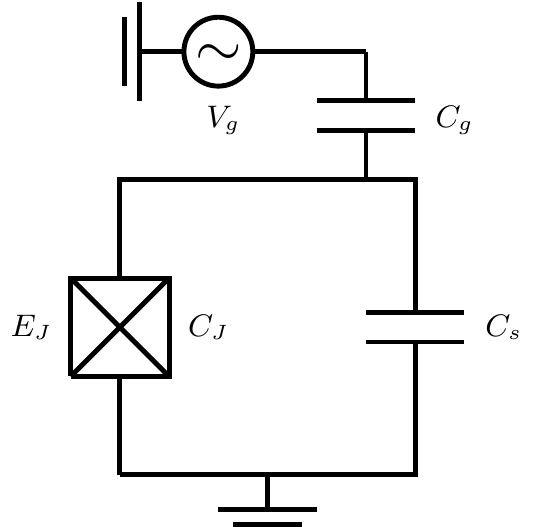}
    % \scalebox{0.95}{\circuitdiagram}
    \caption{Effective circuit diagram of the Cooper pair box. A Josephson junction is placed in parallel to a capacitance $C_s$, and capacitively coupled to a voltage source. In the transmon architecture, this Josephson junction is further shunted by a large secondary capacitance.}
    \label{fig:my_label}
\end{figure}

The design of the transmon qubit \cite{Koch_2007} is similar to the Cooper pair box (CPB) qubit, containing a superconducting island consisting of one or two\footnote{Using one Josephson junction results in a fixed frequency qubit, while the usage of two in a dc superconducting quantum interference device setup allows for the flux tuning of $E_J$. Both cases result in the same Hamiltonian, provided junctions are identical.} Josephson junctions. The transmon qubit is distinguished by a further large capacitance shunting of the Josephson junction, which alters the relative strengths of energy parameters in the system. The system dynamics can be reduced to the isolated qubit CPB Hamiltonian \cite{Koch_2007}
\begin{equation} 
\label{eq:full_ham_no_cav}
\hat{H} = 4E_C(\hat{n} - n_g)^2 - E_J\cos(\hat{\varphi}).
\end{equation} 
Here, $E_J$ and $E_C$ represent the Josephson and charging energies respectively, with $E_C = \frac{e^2}{2C_{\Sigma}}$ where $C_{\Sigma}$ is the effective capacitance $C_J + C_g + C_s$ (see Fig. \ref{fig:my_label}). Any effective offset charge bias in the device is also included as $n_g$. This accounts for unwanted charge noise arising from the environment, or for an external gate voltage $V_g$, used for control purposes. The variables $\hat{n}$ and $\hat{\varphi}$ are the canonical conjugate variables that represent the Cooper pair number and gauge-invariant phase difference operators respectively. 
 
The choice of the ratio $E_J/E_C$ distinguishes different classes of superconducting qubits; in the transmon the Josephson energy dominates the charging energy, resulting in small charge dispersion. This in turn reduces sensitivity to gate charge exponentially with $E_J/E_C$ \cite{Koch_2007}, giving rise to longer coherence times. This is done at the cost of reduced anharmonicity, although it has a much weaker dependence given as $~(E_J/E_C)^{-1/2}$.

For the effective operation of the transmon as a qubit, it is conventional to couple it to a cavity resonator for readout purposes \cite{Koch_2007}. This can be achieved via a gate capacitance. In the event that the gate capacitance is small compared to the transmon and resonator capacitances ($C_g \ll C_r, C_{\Sigma}$), this system is well described by circuit QED, with the effective Hamiltonian ($\hbar$ = 1) written as

\begin{align}
\label{eq:full_ham}
\hat{H}_{\mathrm{CPB}} = &4E_C(\hat{n} - n_g)^2 - E_J\cos(\hat{\varphi}) \\
&+\omega_r \hat{a}^{\dag}\hat{a} - 2e\beta_C V_{rms}^0 \hat{n}(\hat{a}+\hat{a}^{\dag})\nonumber,
\end{align} 
where $\hat{a}$ and $\hat{a}^\dag$ are the canonical lowering and raising operators for the resonator, $\beta_C = C_g/C_{\Sigma}$ is the voltage divider ratio and $V^0_{rms} = \sqrt{\omega_r/2C_r}$. 

In our discussion we take Eq. (\ref{eq:full_ham}) to be the most faithful Hamiltonian, and use it as the baseline model to make further comparisons. While it is possible to solve  Eq. (\ref{eq:full_ham_no_cav}) analytically in terms of Mathieu functions, an analytic solution to the cavity coupled Hamiltonian Eq. (\ref{eq:full_ham}) does not exist; instead, it is solved through numerical diagonalisation. Given the reliance on this numerical approach, it is useful to consider approximations of the underlying model to reduce computational overhead of simulation. In this investigation, we place our attention on three approximations of the underlying model present in the literature; in increasing simplicity, these models will be referred to as the sextic Duffing oscillator ($\mathrm{DO_3}$), the generalised quantum Rabi (GR) model, and the two-level quantum Rabi ($\text{R}$) model. 

The $\mathrm{DO_3}$ model constitutes the first level of approximation and is derived by treating the transmon as a perturbed harmonic oscillator. This approach is justified in the limit of $E_J/E_C \gg 1$, resulting in a weak anharmonicity and conversely, eigenstates with highly localised phase, $\Delta \hat{\varphi} \ll 1$. Consequently the cosine term can be treated as a series expansion truncated at $K^{\mathrm{th}}$ order, 

\begin{equation}
\cos(\hat{\varphi}) \approx \sum_{k=0}^K \frac{(-1)^{k} \hat{\varphi}^{2k}}{(2k)!},
\end{equation}
where at any finite $K$ the periodic boundary conditions are neglected. In the limit of $E_J/E_C \gg 1$, terms beyond quadratic are treated as corrections to the harmonic oscillator. It follows that the number and phase operator can be expressed as
\begin{align}
\label{eq:charge_op_id}
    \hat{n} &= \frac{i}{2\sqrt{\eta}}(\hat{b}^\dag - \hat{b}), \\
    \hat{\varphi} &=  \sqrt{\eta} (\hat{b}^\dag + \hat{b}),
\end{align}
with $\eta = \sqrt{2E_C/E_J}$. For $K > 1$, the uncoupled transmon system can be written in the form of a Duffing oscillator, 
\begin{align}
\label{eq:cos_exp_H}
\hat{H}_{\mathrm{DO}_K} &= \sqrt{8E_CE_J}(\hat{b}^{\dagger}\hat{b} + \frac{1}{2}) - E_J  \nonumber\\
- &E_J \sum_{k=2}^{K} \left(\frac{2E_C}{E_J} \right)^{k/2}\frac{(-1)^k(\hat{b}^{\dag} + \hat{b})^{2k}}{(2k)!},
\end{align}
defining a family of Hamiltonians parametrised by the truncation order $K$.

With this approximation, resonator-induced coupling between transmon states can be justifiably limited to nearest neighbours (see Appendix \ref{appendix:a}). The coupled system Hamiltonian is thus expressed in fundamental transmon literature \cite{Koch_2007} as the canonical Duffing oscillator model,
\begin{align}
\label{eq:sec_ham0}
\hat{H}_{\mathrm{DO_2}} =& \sqrt{8E_CE_J}(\hat{b}^{\dagger}\hat{b} + \frac{1}{2}) - E_J - \frac{E_C}{12}(\hat{b}+\hat{b}^{\dagger})^4 \nonumber \\
&+\omega_r \hat{a}^{\dag} \hat{a}  + ig(\hat{a} +\hat{a}^\dag)(\hat{b}^\dag -\hat{b}).
\end{align}
Note that the prefactors of the asymptotic number operator are absorbed into $g$.

Preliminary analysis in an experimentally relevant $E_J/E_C$ regime with the inclusion of a sextic term ($K=3$) proved to have non-negligible effects on the system spectrum in comparison to the quartic case ($K=2$), causing us to introduce the $\mathrm{DO_3}$ Hamiltonian, 
\begin{align}
\label{eq:sec_ham}
\hat{H}_{\mathrm{DO_3}} =& \sqrt{8E_CE_J}(\hat{b}^{\dagger}\hat{b} + \frac{1}{2}) - E_J - \frac{E_C}{12}(\hat{b}+\hat{b}^{\dagger})^4 \nonumber\\
&+ \frac{E_J}{720}\left(\frac{2E_C}{E_J}\right)^{3/2}(\hat{b}+\hat{b}^{\dagger})^6 \nonumber\\
&+\omega_r \hat{a}^{\dag} \hat{a}  + ig(\hat{a} +\hat{a}^\dag)(\hat{b}^\dag -\hat{b}). 
\end{align}

The GR model can be derived from Eq. (\ref{eq:sec_ham0}) with the inclusion of a perturbative approximation. In the limit $E_J/E_C\gg1$, the relative effect of the sextic term in Eq. (\ref{eq:sec_ham}) becomes negligible and the quartic term can be approximated by first-order eigenenergy corrections. In this case, the Hamiltonian takes the form of a harmonic oscillator with modified eigenenergies for each eigenstate $\ket{m}$, 
\begin{equation}
E_m = -\frac{E_C}{12}(6m^2 + 6m + 3),
\end{equation}
which can be incorporated through the operator,
\begin{equation}
\label{eq:em}
\hat{E}_m = -\frac{E_C}{12}(6(\hat{b}^{\dagger}\hat{b})^2 + 6\hat{b}^{\dagger}\hat{b} + 3).
\end{equation}
This correction yields the GR Hamiltonian
\begin{align}
\label{eq:ham_rwa}
\hat{H}_{\mathrm{GR}} &= \sqrt{8E_CE_J}(\hat{b}^{\dagger}\hat{b} + \frac{1}{2}) - E_J \nonumber\\
              &-\hat{E}_m  + \omega_r \hat{a}^{\dag} \hat{a} + g(\hat{a} +\hat{a}^\dag)(\hat{b}^\dag +\hat{b}),
\end{align}
where the unitary transformation $U = e^{-\frac{i\pi}{2}b^{\dagger}b}$ has been applied to express the coupling term in its conventional form. The key simplification made by this model is the omission of off-diagonal elements in the transmon Hamiltonian. In this picture, the transition frequency between the ground and first excited states can be recovered as $\hbar\omega_{01} \approx \sqrt{8E_CE_J} -E_C$. This model of a transmon system is used within quantum control literature \cite{Zahedinejad_2016, Spiteri_2018, Allen_2017}, often in conjunction with the rotating wave approximation (RWA) \cite{Liebermann_2016, Daraeizadeh_2020, Machnes_2018, Werninghaus_2021, Basilewitsch_2019, Kirchhoff_2018, Wu_2020, Niu_2019}. Usage of the RWA induces negligible error within the parameter regime investigated in this work (and thus is not reported on), but can introduce significant error of its own in regimes of strong coupling or strong driving \cite{Ballester_2012}. The $\text{R}$ model can be obtained directly from the $\text{GR}$ representation by neglecting anharmonic terms, as seen in 
\begin{align}
\label{eq:ham_2qr}
\hat{H}_{\mathrm{R}} &= (\sqrt{8E_CE_J}-E_C)\frac{\hat{\sigma}_z}{2} \nonumber\\
              &+ \omega_r \hat{a}^{\dag} \hat{a} + g\hat{\sigma}_x(\hat{a}+\hat{a}^{\dagger}),
\end{align}
where the $\sigma_i$ denote the Pauli matrices. Omitting higher energy levels is a negligent approach to transmon simulation. As such, the use of this model is generally restricted to research that focuses on innovation within the control algorithm at the expense of modelling a specific qubit architecture \cite{Zhang2019, An_2019, Bao_2018}.

\subsection{Control}
Transmon control can be realised through a capacitively coupled time-dependent driving voltage $V_g(t)$. The driving Hamiltonian of interest is
\begin{equation}\label{eq:driv}
   \hat{H}_d(t) =  \hat{n}V_g(t).
\end{equation}
For models operating in the anharmonic oscillator basis ($\hat{H}_{\mathrm{DO_3}}$, $\hat{H}_{\mathrm{GR}}$, and $\hat{H}_{\mathrm{R}}$), Eq. (\ref{eq:driv}) can be re-expressed in terms of the operator introduced in Eq. (\ref{eq:charge_op_id}),
\begin{equation}\label{eq:drive_ham}
     \hat{H}_d =  \frac{i}{2\sqrt{\eta}}(\hat{b}^\dag - \hat{b})V_g(t).
\end{equation}
We format the drive pulse into a general sum of sinusoidal carrier frequencies modulated by an envelope function,
\begin{equation}\label{eq:drive_ham_potential}
    V_g(t) = \sum_{i} \Omega_{i}(t)\cos(\omega_i^{dr} t + \gamma_{i}),
\end{equation}
where $\omega_i^{dr}$ is the driving frequency, $\gamma_{i}$ is the offset phase and $\Omega_{i}$ is the envelope function. In experimental situations, the amplitude of the envelope function should be constrained to zero at $t=0$ and $t=T$, with $T$ representing the final time of the pulse. Additional constraints are also introduced by the sampling rate of the hardware, which constrains the minimum time under which an arbitrary envelope function may be expressed.

Information leakage outside the computational subspace is a predominant issue in transmon control. A common technique to combat leakage is to apply a two-component drive pulse expressed as
\begin{equation}
    V_g(t) = \Omega(t)\cos(\omega^{dr} t) -\beta \dot{\Omega}(t) \sin(\omega^{dr} t),
\end{equation}
with parameter $\beta$ representing a coefficient that maximally suppresses leakage when tuned. This technique is known as derivative removal by adiabatic gate (DRAG) \cite{motzoi_gambetta_rebentrost_wilhelm_2009}. 
\

\section{\label{sec:Methods} Methods}
We have presented four models of a transmon qubit architecture. Taking the CPB Hamiltonian $\hat{H}_{\mathrm{CPB}}$ as the baseline description of the system, $\hat{H}_{\mathrm{DO_3}}$, $\hat{H}_{\mathrm{GR}}$, and $\hat{H}_{\mathrm{R}}$ are approximations with decreasing levels of sophistication. As detailed in Sec. \ref{sec:Theory}, each model is approximately equivalent in the regime $E_J/E_C \gg 1$. Despite the prevalence of these approximations in the literature, a comparative study of the associated models with experimentally relevant parameters is lacking (although $\hat{H}_{\mathrm{CPB}}$ has been well examined \cite{Willsch2017}). We seek to explore this area within the context of quantum control, where correspondence between results of simulated protocols and protocols run on hardware is of clear importance.

We initially explore the differences in behaviour evident in the spectrum of each model, specifically the parameters relevant to quantum control: qubit frequency and anharmonicity. The extent of spectral deviation observed between individual models provides preliminary insight into the level of approximation sufficient for simulation of a real transmon device.

As we are concerned with the validity of these assumptions in the context of engineering high-fidelity pulse sequences, the dynamics of each model will be evaluated when probed by identical Gaussian pulses. Each model will then be used to design an optimal Gaussian $\frac{\pi}{2}$-pulse, by finely sweeping over amplitudes for Gaussian pulses of constant timespan and observing $\ket{0}$ and $\ket{1}$ populations. In order to generate pulses for deployment on an experimentally relevant system, we match our simulation environment parameters to that of a real experimental device, for which purpose we select the remotely available \textsf{IBMQ} device \texttt{ibmq\_armonk}. Device parameters are obtained by a series of experiments conducted on the \texttt{ibmq\_armonk} system; details of these experiments are available in Appendix \ref{appendix:b}. Generated $\frac{\pi}{2}$-pulses will then be deployed on \texttt{ibmq\_armonk} and evaluated by randomised benchmarking procedures to establish gate fidelities \cite{Magesan_2011}. This will subsequently allow us to quantify the losses when transferring to hardware introduced by each level of model simplification. 

In practice, the simplicity of the control parametrisation used in the above experiment means that conventional amplitude scaling and frequency calibration in laboratory conditions have the capacity to mitigate error induced by use of a simplified model. We emphasise that these methods cannot be used in general to correct approximation-induced errors. To do this, physical origins of the differences between model Hamiltonians are described and numerical simulations are performed to demonstrate the importance of the pulse frequency profile. 

\begin{figure*}[!htbp]

\begin{subfigure}{.5\textwidth}
  \centering
  \includegraphics[width=0.93\linewidth]{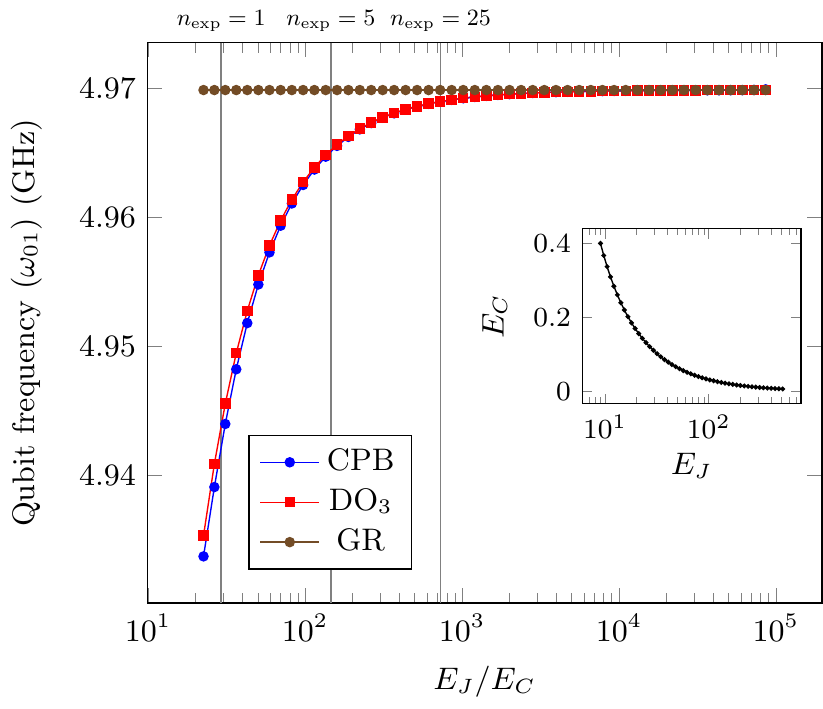}
%   \scalebox{1}{\EjcFreqInset}
  \caption{Qubit frequency and $E_J/E_C$}
  \label{fig:EjcFreqInset}
\end{subfigure}%
\begin{subfigure}{.5\textwidth}
  \centering
  \includegraphics[width=0.93\linewidth]{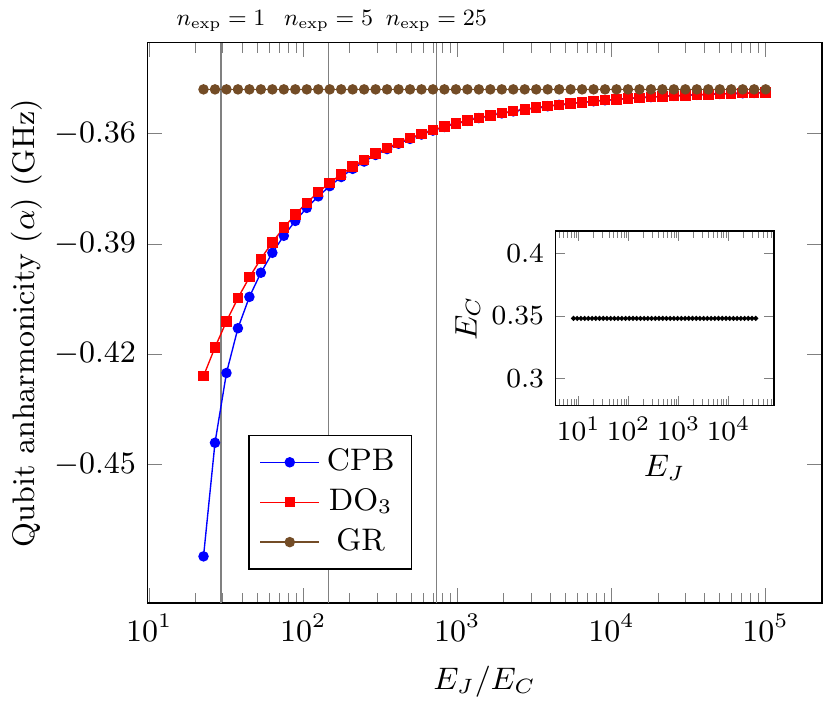}
%   \scalebox{1}{\EjcAnharmInset}
  \caption{Anharmonicity and $E_J/E_C$}
  \label{fig:EjcAnharmInset}
\end{subfigure}
\caption{Spectral features of a coupled transmon-resonator system, under a range of modelling schemes and $E_J/E_C$ ratios. Panel (a) demonstrates significant qubit frequency deviations between modelling approaches for an experimentally relevant regime ($n_{\mathrm{exp}} = 1$) with $\omega_{01}$ in the GR model held constant, while (b) demonstrates further deviations in the qubit anharmonicity across models with $E_C$ (equivalent to $\alpha$ in the GR model) held constant. In each case, the model spectra are seen to converge as $E_J/E_C \rightarrow \infty$. Inset graphs illustrate how $E_C$ and $E_J$ were individually varied to increase the $E_J/E_C$ ratio; in each case, optimised to maintain the GR spectral feature of interest.}
\label{fig:EjcInsets}
\end{figure*}

To extend this investigation, we finally seek to quantify the difficulty of reaching an optimal control solution in each model, rather than contrasting the solutions themselves. To this end, the Gaussian control landscape evaluated by first excited level population and parametrised by pulse time and amplitude will be compared for each model. In order to produce these landscapes, the evolution of a ground-state transmon-resonator system is simulated (with each of the four modelling approaches) over a sweep of each control parameter.

The landscape investigation will then be extended to a DRAG pulse parametrisation. The gradient optimisation technique GOAT ~\cite{Machnes_2018} will be used to generate a set of control trajectories, iteratively determining an optimal Gaussian amplitude and tunable DRAG coefficient from a distribution of starting points. With this tool, we seek to quantify the complexity of navigation in a given control landscape more stringently, introducing the metric $R_\gamma$ of a trajectory $\gamma$ as \cite{Nanduri_2013}
\begin{align}\label{equ:rmetric}
    R_{\gamma} = \frac{d_{\gamma}}{d_{E_{\gamma}}},
\end{align}

where $d_{\gamma}$ and $d_{E_\gamma}$ are the path length of $\gamma$ and the Euclidean distance between the initial and final points of $\gamma$ respectively. As $d_{\gamma} \rightarrow d_{E_\gamma}$, the lower bound of the $R_{\gamma}$-metric saturates and the control trajectory $\gamma$ is considered optimal. Insight into the structural complexity can be established by averaging the $R_{\gamma}$ metric over a set of gradient-optimised control trajectories $\Gamma$. The distribution of starting points for these control trajectories is chosen to be a multivariate Gaussian distribution centred around an established optimum.

\section{\label{sec:Contrasting_descriptions} Results}
\subsection{System characterisation \label{sub:char_results}}

In order to produce results relevant to current transmon-resonator systems, system parameters residing within order-of-magnitude bounds of publicly accessible \textsf{IBMQ} devices are selected, and expressed in Table \ref{tab:ArmonkParams}.

\begin{table}[!htbp]
\begin{centering}
{\renewcommand{\arraystretch}{1.2}
\begin{tabular}{@{}cccc@{}}

\toprule
\hline
\hline
$E_C/2\pi$ (GHz) & $E_J/2\pi$ (GHz) & $g/2\pi$ (GHz) & $\omega_r/2\pi$ (GHz) \\ 
\midrule
\hline
0.348 & 10.158 & 0.02 & 6.99 \\ 
\hline
\hline
\bottomrule
\end{tabular}}
\end{centering}
\caption{Energy parameters for transmon-resonator simulation. \label{tab:ArmonkParams}}
\end{table}

As is common in experimental transmon literature, the ratio of $E_J$/$E_C$ is of the $10^1-10^2$ order of magnitude. The frequent usage of the assumption $E_J/E_C \gg 1$ in model derivations motivates investigation into the effects of this energy scaling on system properties. Figure \ref{fig:EjcInsets} illustrates the dependence of qubit frequency and anharmonicity on the magnitude of $E_J/E_C$ for each model. Given the two degrees of freedom that one can use to increase the ratio of $E_J/E_C$, we opt to perform this operation in two distinct ways in Fig. \ref{fig:EjcInsets}: (a) constant Rabi model frequency $\omega_{01} = \sqrt{8E_JE_C} - E_C$ and (b) constant Rabi model anharmonicity $\alpha = -E_C$. Taking the previously defined simulation parameters, $E_J^{\mathrm{exp}}/2\pi = 10.158 $ GHz and $E_C^{\mathrm{exp}}/2\pi = 0.348 $ GHz, we define the parameter $n_{\mathrm{exp}}$ to be the multiplier of the standard experimental ratio  
\begin{align}
    \frac{E_J}{E_C}  =  \left(\frac{E_J^{\mathrm{exp}}}{E_C^{\mathrm{exp}}}\right)n_{\mathrm{exp}} = 29.19\:n_{\mathrm{exp}}.
    \end{align}

\begin{figure}
    \centering
    \includegraphics[width=0.93\linewidth]{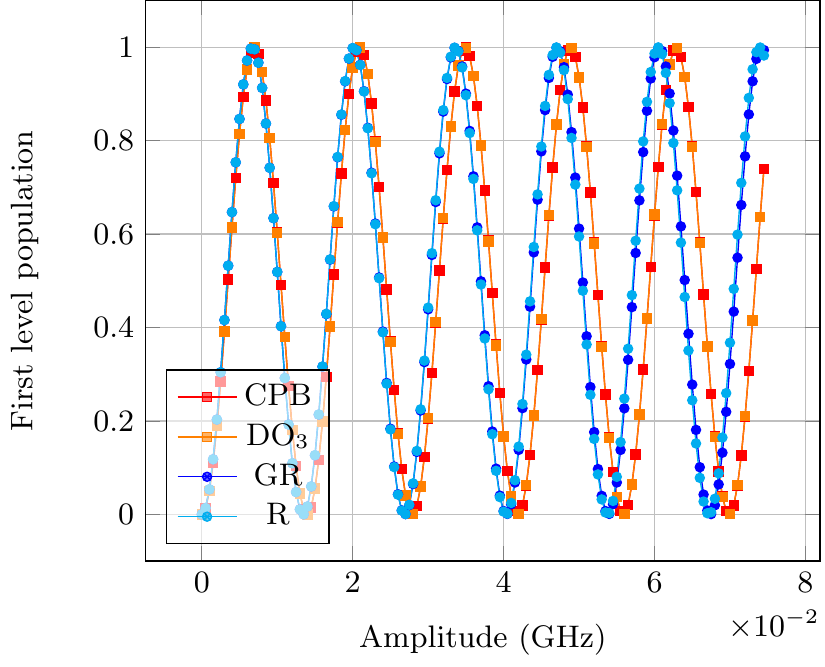}
    % \scalebox{0.95}{\EjcIPop}
    \caption{The dynamics induced in each model by a Gaussian pulse with period 142.2 ns, over a set of pulse amplitudes from 0 to 75 MHz. The first excited level population is taken to be representative of the dynamics in each case. There is a clear divergence in population between the family of models in red (CPB) and orange ($\mathrm{DO_3}$), and the family of models in dark blue (GR) and light blue (R). This result implies that the efficacy of optimised control fields may depend markedly on the model used in the optimisation process.}

  \label{fig:EjcIPop}
\end{figure}

Figure \ref{fig:EjcInsets} demonstrates significant spectral variance across models, with qubit frequency and anharmonicity deviating from CPB to GR by over 15 and 50 MHz, respectively, at $n_{\mathrm{exp}} = 1$. To establish whether these model inconsistencies affect system dynamics, each simulated system is probed by an identical Gaussian pulse mixed with a carrier signal at their respective frequencies. A pulse period of 142.2 ns is used, with amplitudes ranging from 0--75 MHz.

Examining Fig. \ref{fig:EjcIPop}, the population of the first excited level is plotted against the amplitude of the probing Gaussian pulse for each of the four models in the regime $n_{\mathrm{exp}}=1$. All are observed to exhibit periodic behaviour interpreted as Rabi oscillations. As the amplitude is increased, the influence of approximations on the dynamics of each system becomes apparent. 

It can be observed that the first-order perturbative correction, employed for the GR and R models but absent for the CPB and $\mathrm{DO_3}$ models, induces a pronounced deviation in dynamics at later Rabi periods due to a mismatch of Rabi frequencies. Appendix \ref{appendix:b.5} presents an auxiliary result that verifies that this difference is predominantly attributable to the first-order perturbative correction rather than the neglection of the sextic term, both of which occur in the transition from $\mathrm{DO_3}$ to GR.
Meanwhile, the omission of effects from outside the computational subspace (comparing GR and R) leads to a small but perceptible variance in dynamics at high drive amplitudes, while the basis change, cosine expansion and truncation implicit in the comparison of CPB and $\mathrm{DO_3}$ have no apparent effect. 

\subsection{Validation on hardware \label{hardware}}

To verify that the hierarchy of modelling approaches presented corresponds to decreasing fidelity with a real system, and to examine the degree of these differences, a validation procedure on hardware (\texttt{ibmq\_armonk}) is conducted. To do this, single qubit gates are designed in simulation for each model and subsequently evaluated by a random benchmarking protocol on hardware.

In order to characterise the \texttt{ibmq\_armonk} device hardware and design the appropriate gates, a set of standard calibration experiments are performed to approximate system parameters. These calibrations are detailed in Appendix \ref{appendix:b}. Energy parameters in simulation are then replaced with the values obtained via these experiments.

\begin{figure*}
\begin{subfigure}{.5\textwidth}
  \centering
  \includegraphics[width=0.93\linewidth]{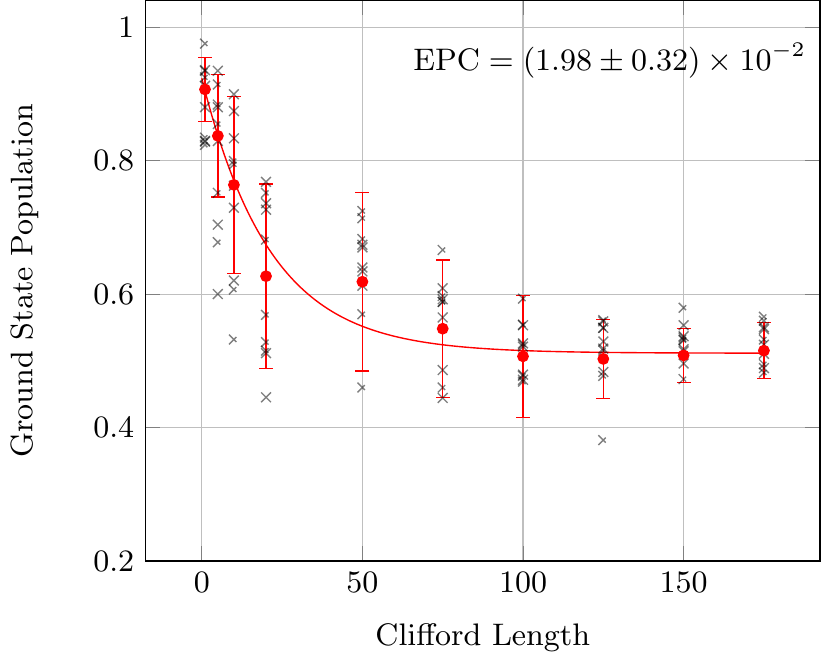}
%   \scalebox{.9}{\RandomBenchCPB}
  \label{fig:rb_cpb}
\end{subfigure}%
\begin{subfigure}{.5\textwidth}
  \centering
  \includegraphics[width=0.93\linewidth]{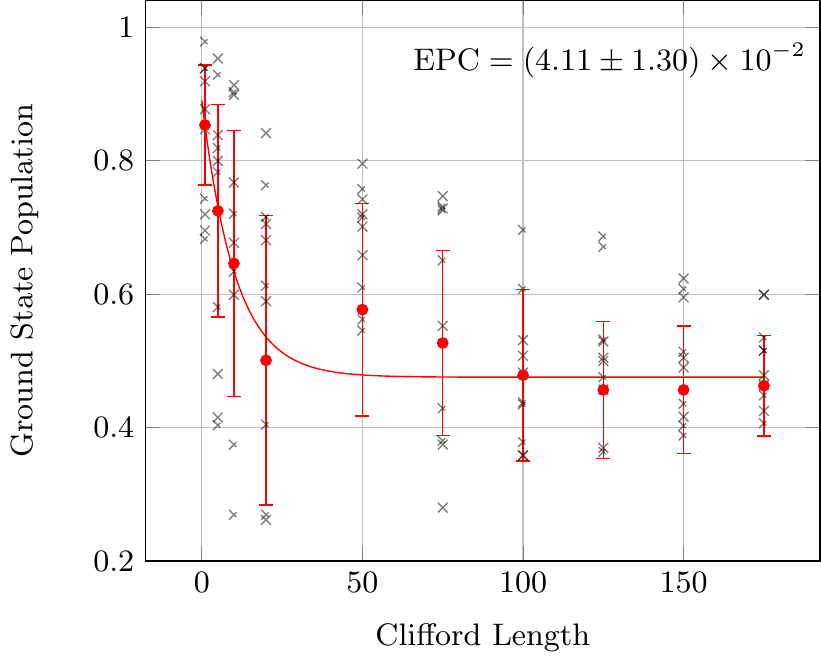}
%   \scalebox{.9}{\RandomBenchRWA}
  \label{fig:rb_qr}
\end{subfigure}
\caption{Randomised benchmarking results for a single-qubit Clifford gate set, transpiled using RX($\frac{\pi}{2}$) pulses. The pulse parameters used in (a) and (b) were optimised using $\mathrm{DO_3}$ and GR models respectively. The EPC (error per Clifford gate), established by exponential curve fitting, increases by over a factor of 2 from $\mathrm{DO_3}$ to GR. This result substantiates the earlier inference that the efficacy of an optimised control field depends markedly on the model used in the optimisation process.}
\label{fig:randbench}
\end{figure*}

To universally perform any single-qubit gate, only RZ($\theta$) and RX($\frac{\pi}{2}$) pulses require definition. To see this, consider the single-qubit universal rotation gate

\begin{align} U3(\theta, \phi, \lambda) &= \begin{pmatrix} \cos{\frac{\theta}{2}} & -e^{i\lambda}\sin{\frac{\theta}{2}} \\ e^{i\phi}\sin{\frac{\theta}{2}} & e^{i(\phi+\lambda)}\cos{\frac{\theta}{2}} \end{pmatrix} \nonumber \\
&= RZ(\phi)RX(-\frac{\pi}{2})RZ(\theta)RX(\frac{\pi}{2})RZ(\lambda).
\label{eq:transpiler}
\end{align}

The Qiskit application programming interface (API) provides optimised pulses for these two gates, implementing RZ($\theta$) as a virtually executed change of frame and RX($\frac{\pi}{2}$) as a 142.2 ns pulse. To evaluate the accuracy of each proposed transmon-resonator model, we seek to replace the default RX gate with a Gaussian pulse, described by optimal parameters established within the corresponding simulation environment. In each case, the pulse period is fixed (142.2 ns) and amplitude is optimised to produce the RX($\frac{\pi}{2}$) gate of highest fidelity. The Qiskit-optimised default pulse is then replaced for transpilation by the numerically optimised pulse, and randomised benchmarking sequences are subsequently transpiled and evaluated with the single-qubit Clifford gate set \{H, X, Y, Z, S, S$^{\dagger}$\}.

The numerical optimisation (via amplitude sweeping) of RX($\frac{\pi}{2}$) pulses over 142.2 ns leads to amplitudes of 3.38 MHz for the CPB and $\mathrm{DO_3}$ models, whilst amplitudes of 3.25 MHz are obtained for the GR and R models, as seen in Table \ref{tab:rbresults}. Randomised benchmarking sequences conducted with CPB- and $\mathrm{DO_3}$-optimised pulses produce error per Clifford metrics of $1.98\times10^{-2}$ in application to hardware. Corresponding sequences conducted with GR- and R-optimised pulses produce error per Clifford metrics of $4.11\times10^{-2}$, an error magnification of over 100\% (as seen in Fig. \ref{fig:randbench}). This substantiates the claim that simulated control protocols optimised using Rabi models will yield inflated error rates when directly migrated to hardware. As will be addressed in the upcoming control landscape results, this divergence is likely to be amplified under a stronger drive.

\begin{table}[!htbp]
\begin{centering}
{\renewcommand{\arraystretch}{1.4}
\begin{tabular}{@{}c|c|c|c|c@{}}

\toprule
\hline
\hline
 & \textbf{$\hat{H}_{\mathrm{CPB}}$} & \textbf{$\hat{H}_{\mathrm{DO_3}}$} & \textbf{$\hat{H}_{\mathrm{GR}}$} & \textbf{$\hat{H}_{\mathrm{R}}$} \\ 
\midrule
\hline
Amplitude (MHz) & \multicolumn{2}{c|}{3.38} & \multicolumn{2}{c}{3.25} \\
\hline
EPC & \multicolumn{2}{c|}{$1.98\times10^{-2}$} & \multicolumn{2}{c}{$4.11\times10^{-2}$} \\ 
\hline
\hline
\bottomrule
\end{tabular}}
\end{centering}
\caption{Optimised amplitudes for a 142.2 ns RX($\frac{\pi}{2}$) Gaussian pulse, and error per Clifford (EPC) associated with each as evaluated by randomised benchmarking on \texttt{ibmq\_armonk}. \label{tab:rbresults}}
\end{table}

The error per Clifford metric approaches or exceeds $2\times10^{-2}$ in all cases. Improvements in this metric are available through two major areas. The first is a refined model, through modelling noise and dissipation, and/or obtaining access to precise system parameters, thereby removing reliance on calibration experiments. A second pressing area is control pulse enhancements; specifically, the usage of more sophisticated pulse parametrisations, or a more efficient Clifford gate transpilation. This transpilation would be achieved by using a basis set of fundamental pulses that can compile Clifford gates using less gates on average than that of Eq. \eqref{eq:transpiler}. Despite this scope in improvement with respect to absolute fidelity, it is clear that the first-order perturbation applied when moving to the Rabi family of models incurs a damaging cost in simulation precision even for simple single-qubit controls.

It is important to note that in a laboratory environment, the calibration of drive amplitudes for single-qubit control occurs with the system in-the-loop. Indeed, the error type described above can be largely mitigated by a simple drive amplitude calibration across models. 
% Without further investigation into the emergence of these errors, and the consequences of their existence, this demonstration of imprecision is not deeply concerning. The purpose of this experiment on hardware is to validate the operational premise which underpins this investigation; there exists a marked decrease in fidelity when moving from a fundamental system model to a generalised Rabi model for simple driven transmon-based architectures. With this demonstration of fidelity decay in hand, 
We now seek to characterise a set of control settings in which errors caused by these approximations cannot be mitigated in experimental conditions.

\subsection{Detuned control \label{sec:origins}}

Throughout this section, experiment-like parameter calibrations are applied within simulation to maximally match DO$_3$ and GR dynamics in the computational subspace. Specifically, we apply an amplitude scaling factor to each applied pulse, calibrated such that Rabi oscillations are matched between each model (mitigating the error seen in Fig. \ref{fig:EjcIPop}). Subsequently, we vary drive frequency (for each pulse strength used) and observe the periodicity of Rabi oscillations to approximate the Stark-shifted frequency associated with each (note that differences in Stark shift are trivial). We perform these steps to emphasise that the model-induced deviations are in general not resolvable through simple parameter calibrations, and originate from the omission of off-diagonal Hamiltonian elements and overall mistreatment of dynamics in the noncomputational subspace.

The differences between the compositions of $\hat{H}_{\mathrm{DO}_3}$ and $\hat{H}_{\mathrm{GR}}$ are minimal. Energy parameters E$_J$ and E$_C$ can be tuned to match qubit frequency and anharmonicity across transmon Hamiltonians, and the drive Hamiltonians are by definition equivalent. Under these conditions, the only material differences between the models are frequencies of higher-level transitions such as $\ket{2}\rightarrow\ket{3}$ and so on, and the existence of off-diagonal elements in the Hamiltonian for $\hat{H}_{\mathrm{DO}_3}$ that provide couplings between distant energy states such as $\ket{0}\rightarrow\ket{2}$ and $\ket{1}\rightarrow\ket{3}$. 

\begin{figure}
\centering
    \includegraphics[width=0.93\linewidth]{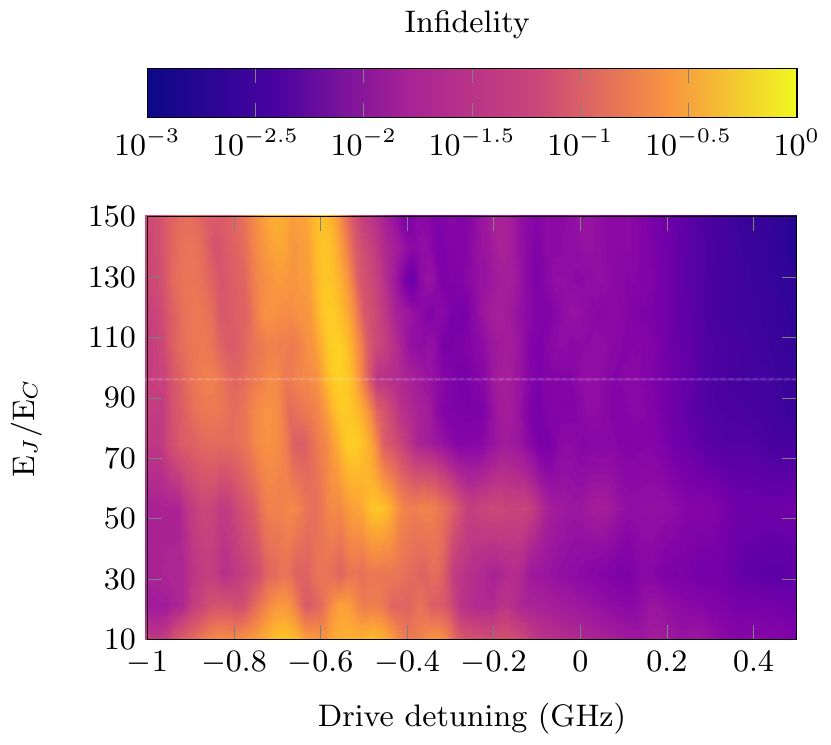}
    % \scalebox{0.96}{\FreqResponseDeviations}
    \caption{Deviations in simulated control between the DO$_3$ and GR models. The $x$ axis is detuning of a 5 ns drive pulse from the qubit frequency. The $y$ axis is $E_J/E_C$ values for defining the GR Hamiltonian. $E_J$ and $E_C$ values of a comparable ratio were chosen to precisely match the qubit frequency and anharmonicity for the DO$_3$ Hamiltonian. Amplitudes around 190 MHz were chosen for the drive, scaled to maximally match dynamics. Both models undergo evolution via 5 ns square pulses. The $z$ axis represents the maximum infidelity experienced between models from a subset of times during the evolution of the 5 ns square pulse, beginning in a computational superposition state. As can be seen, maximum infidelities lie within the range of frequencies associated with higher-level transitions, and exist across all $E_J/E_C$ values. Infidelity is artificially increased at around $E_J/E_C \sim 50$ due to the qubit frequency briefly nearing the resonator frequency.}
    \label{fig:freq_response}
\end{figure}

The extent to which leakage out of the computational subspace becomes problematic for different energy regimes is illustrated in Fig. \ref{fig:freq_response}. Deviations in predicted leakage fluctuate with two key properties; decreases in the $E_J/E_C$ ratio, which is previously established as a measure of model divergence, and the frequency profile of the drive pulse, which determines the activations of energy level transitions and thereby the intrinsic model differences in noncomputational dynamics. There exist clear infidelity maxima in the negative drive detuning region. This corresponds to the range of drive frequencies associated with higher-level and two-photon transitions. To confirm that it is these transitions that give rise to the infidelities observed, a cross-section of the plot taken at the indicative white line is presented in Appendix \ref{appendix:c.5}, with relevant higher-level transitions and two-photon transitions characterised.

The inability to mitigate these errors through control parameter calibration indicates that these are the major deviations of concern in practice. Algorithmic searches for optimal quantum control schemes, in particular those facilitated by neural network-based algorithms, yield pulse shapes that have unintuitive profiles in both the frequency and time domains. It is evident in Fig. \ref{fig:freq_response} that designing any control pulse with non-negligible frequency components within a sub-GHz detuning range relative to the qubit frequency will risk significant and irreversible fidelity loss, without a measured consideration of the model in use. This danger is most pertinent for unorthodox tailored pulse shapes, but should also be taken into consideration in simulation of existing gate implementations that exploit the noncomputational subspace \cite{Chow2013, Magesan2020, Negrneac2021}. 

A set of recommendations are synthesised for control practitioners as follows.

\subsubsection{Resonant pulse design}

The term resonant is used here to denote pulse shapes tightly distributed around the target transition frequency (e.g. low amplitude Gaussians). For a single-qubit or few-qubit experiment operating within this control scheme, it is likely that the deviations demonstrated in this work will have little functional importance. The nature of the induced errors in this regime and the small quantity of control parameters means that calibration in the loop is feasible, expressly that pulse shapes can be straightforwardly scaled and frequency-shifted to achieve maximal fidelities.

As systems scale up to many-qubit algorithms, the full characterisation of individual components will become increasingly prohibitive and simulated spectral predictions for leakage transitions will become an important factor. Spectral crowding within neighbouring qubits (directly resulting in crosstalk error) means that an accurate knowledge of leakage transitions within the frequency domain becomes crucial to the design of control protocols. It is clear in Fig. \ref{fig:freq_response} that even with qubit frequency and anharmonicity values in the GR model scaled to match those in the DO$_3$ model, the dynamical disagreements around leakage transitions are substantial enough to prohibit its use for such a task.

We do not recommend great care when selecting a simulation model for resonant single-qubit control protocols. In contrast, designing a protocol that successfully navigates a crowded spectrum (e.g. many-qubit algorithms) while minimising the activation of undesired transitions forbids the use of the first-order perturbation approximation intrinsic to the general Rabi model, on account of the infidelities observed in Fig. \ref{fig:freq_response}. This is a problem that plagues this class of models throughout the range of experimentally relevant E$_J$/E$_C$ ratios.

\subsubsection{Advanced pulse design}
In the endeavour to increase overall operational fidelity of near-term superconducting systems, high-dimensional multiparameter optimisation tools for discovering efficient pulse protocols has blossomed into a popular field of research \cite{Niu_2019, Bukov_2018, Wu_2019}. It is well established that a) the bulk of this optimisation is often performed in simulation in order to facilitate large iteration volumes and b) the pulses developed have nonstandard profiles in both the time and frequency domains, which they are able to exploit to increase gate fidelities. In addition to algorithmically designed controls, there are existing gate implementations which make use of higher levels and `exotic' transitions such as two-photon transitions. It is these two audiences which we aim to reach principally.

In this case, the recommendations are clear. Algorithmic approaches and exotic gate implementations rely on the mitigation or exploitation of leakage transitions, and to maximise the power and applicability of your protocol it is crucial that the dynamics in these frequency regimes are modelled accurately. We expect that the use of explicit higher-order nonlinear terms in the transmon Hamiltonian (such as in $\hat{H}_{\mathrm{CPB}}$ and $\hat{H}_{\mathrm{DO}_3}$) will help to avoid losses as simulated in Fig. \ref{fig:freq_response} at all experimentally relevant E$_J$/E$_C$ ratios.

\subsection{Control landscapes}
With our attention so far on understanding the regimes within which different models will provide meaningfully different globally optimal control protocols, we now seek to quantify an objective difficulty of determining this solution within each model. In short, we seek to synthesise and demonstrate a method of measuring `ease of control'. A general characterisation of this problem is provided by quantum control landscapes, which are the subject of active research \cite{Chakrabarti2007, Moore_2008, Larocca_2018, Day_2019, Kosut2019}. These structures are defined as
% manifold defined by the loss function associated with a set of control variables
the manifold generated by mapping between control Hamiltonian parameters and some control performance metric. A thorough analysis of control landscapes requires an examination of both the topology (local and global optima) and nontopological features (local structure) of the manifold. Pairing these considerations will provide a basis from which the difficulty of discovering optimal control protocols in each regime can be assessed. 

\begin{figure*}
    \begin{subfigure}[t]{0.49\textwidth}
    \centering
    \includegraphics[width=1\linewidth]{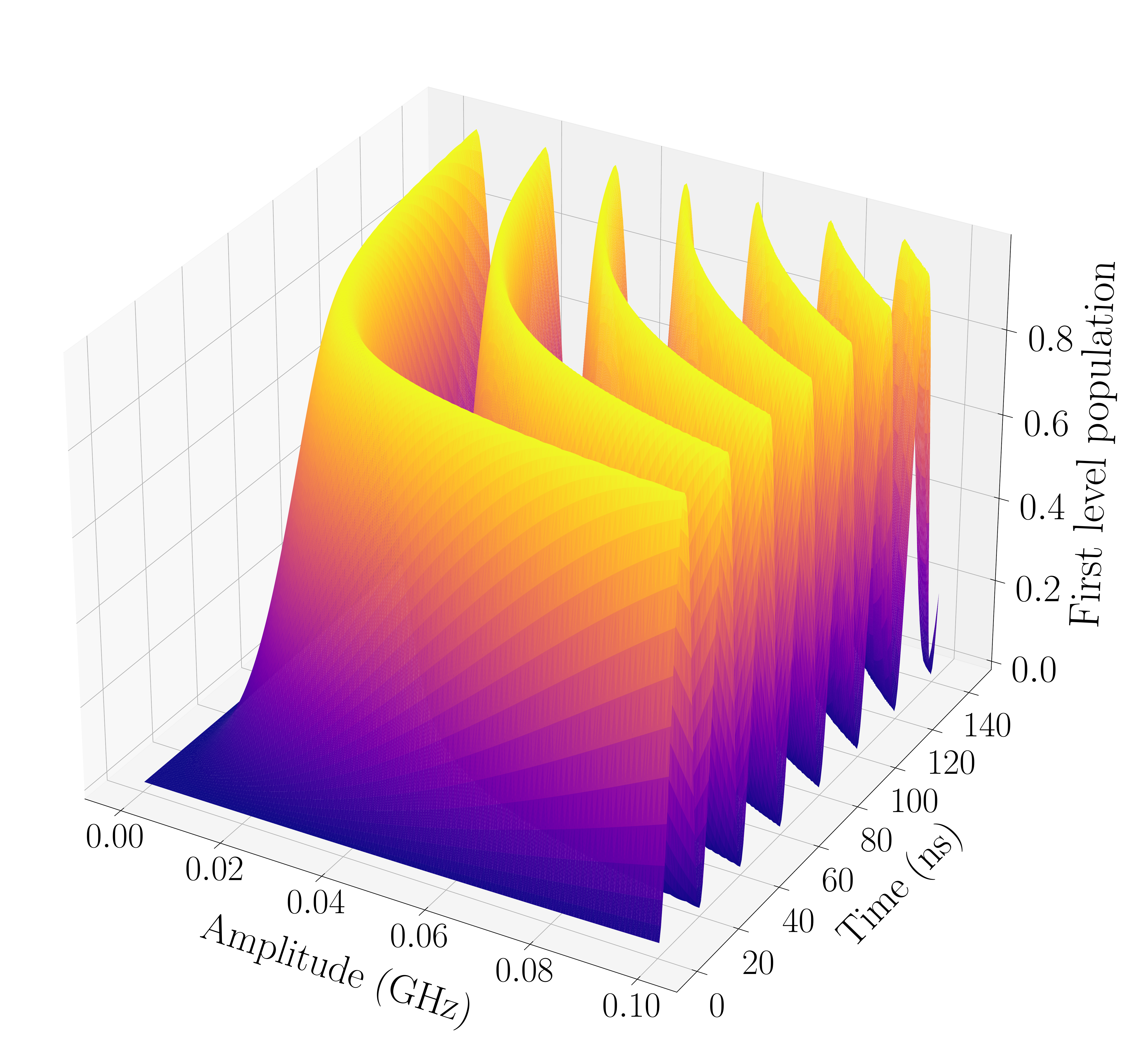}
    \caption{Gaussian control landscape ($\hat{H}_{\mathrm{CPB}}$)}
    \label{fig:CPBlandscape}
    \end{subfigure}
    \begin{subfigure}[t]{0.49\textwidth}
    \centering
    \includegraphics[width=0.93\linewidth]{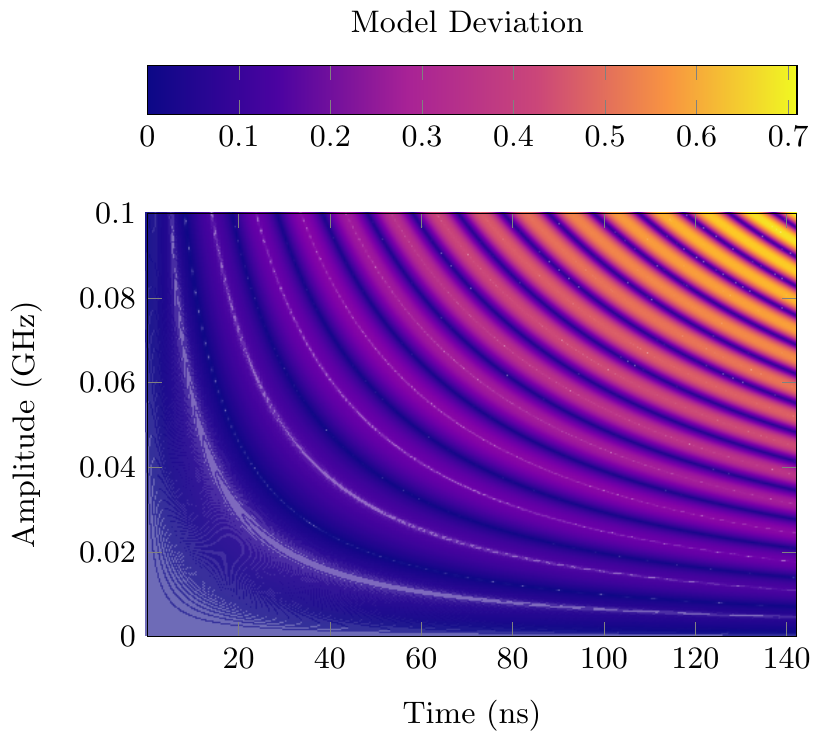}
    % \scalebox{0.8}{\ContourRWATenlvls}
    \caption{Deviations in Gaussian control landscape ($\lvert\hat{H}_{\mathrm{DO_3}} - \hat{H}_{\mathrm{GR}}\rvert$)}
    \label{fig:landscapeDiff}
    \end{subfigure} \hfill
    \caption{(a) The control landscape of the CPB model under application of a Gaussian control pulse. The $z$ axis shows the population of the coupled system state associated with computational $\ket{1}$, after being driven by a Gaussian control pulse with parameters defined on the $x$ and $y$ axes. (b) A visualisation of the deviations in simulated control between the $\mathrm{DO_3}$ and GR models. The axes represent pulse parameters in the same way as (a). The colour intensity of the figure is then the absolute difference in first-level populations after application of the relevant Gaussian control pulse. This figure illustrates the magnitude of model deviation that can occur for high-energy control fields.}
    \label{fig:landscapes}
\end{figure*}

We realise the landscape objective function as the population of the computational $\ket{1}$ state, to act as a probe of the system dynamics. As a preliminary investigation, landscapes are constructed in a parameter space defined by the amplitude and timespan of the Gaussian control pulse. Across all models similar topological features are present, with landscapes punctuated by ridges of high population interpreted as Rabi oscillations. This basic phenomenon can be observed in Fig. \ref{fig:CPBlandscape}, which presents the control landscape under implementation of $\hat{H}_{\mathrm{CPB}}$. In the case of the $\hat{H}_{\mathrm{R}}$ model, these ridges possess a degenerate structure and consequently globally optimal protocols are accessible to naive gradient-based optimisation algorithms. As for the remaining models, the first excited level population is increased by travelling along a ridge in the direction of decreasing pulse amplitude, conducive to reducing leakage. Contrasting globally optimal points across the models  $\hat{H}_{\mathrm{CPB}}$,  $\hat{H}_{\mathrm{DO_3}}$ and  $\hat{H}_{\mathrm{GR}}$, we find that they correspond to different ridges (Rabi oscillations). This implies that not only will the initial conditions of a gradient-based optimisation determine whether the process will achieve a globally optimal solution, but the optimal initial conditions also vary across each model. 

In Fig. \ref{fig:landscapeDiff}, the absolute difference between control landscapes for the $\mathrm{DO_3}$ and GR models are displayed. These models are chosen for contrast due to the earlier identification that the transition between them encompassed the most disruptive assumption (first-order perturbation correction with $E_J/E_C \gg 1$). This figure, which demonstrates excited state population deviations in excess of 70\% within a realistic pulse parametrisation space, acts as an extension of Fig. \ref{fig:EjcIPop} in illustrating the inaccuracies inherent in the generalised quantum Rabi model, although we again stress that for this simple control parametrisation the error can be mitigated by simple amplitude scaling.

Additionally, Fig. \ref{fig:landscapeDiff} lends further legitimacy to the randomised benchmarking results presented in Fig. \ref{fig:randbench}. Due to drive amplitude limits imposed on the \texttt{ibmq\_armonk} device used for benchmarking, the final optimised gates had strengths of just $\sim$3 MHz. Locating this drive within the parameter space in Fig. \ref{fig:landscapeDiff} sees a model deviation of just a few percent, indicative of the fact that this extreme low-amplitude regime produces minimal model deviations in comparison to other regions in parameter space. It can be extrapolated that the deviations observed in randomised benchmarking results would be intensified considerably on hardware without these drive amplitude constraints. 

With a simple Gaussian control protocol, we have seen that the modelling approach impacts both the distribution of global optima and their corresponding fidelity markedly. Elaborating on this initial analysis, we seek to examine the control landscape of a DRAG pulse protocol. This parametrisation removes the triviality of finding local optima in the Gaussian control space (methodology of minimising amplitude at the cost of pulse time) in favour of well-defined optima which maximally suppress leakage. Upon identifying these optima, $10^3$ starting points are randomly sampled from a Gaussian distribution around each. The gradient-based pulse optimisation algorithm GOAT is then used to generate optimal control trajectories for each point. In each trajectory, termination is achieved by meeting one of two conditions: (1) first excited level population exceeding $1\!-\!(5\times10^{-5})$, or (2) trajectory steps exceeding 100. Generally, trajectories are terminated by condition (1). Terminations that occurred by condition (2), but had their final point in a position where other trajectories had successfully converged from, are assumed to converge along the path of other said trajectories.  In Table \ref{tab:RMetric}, the distribution of $R_\gamma$ metrics [defined in Eq. (\ref{equ:rmetric})] over the set of control trajectories generated by GOAT is presented for each model. The premise of the $R_\gamma$ metrics is detailed in Sec. III, whereby the metric tends towards unity as trajectories approach linearity.

\begin{table}[!htbp]
\begin{centering}
{\renewcommand{\arraystretch}{1.4}
\begin{tabular}{@{}cccc@{}}

\toprule
\hline
\hline
\textbf{$\hat{H}_{\mathrm{CPB}}$} & \textbf{$\hat{H}_{\mathrm{DO_3}}$} & \textbf{$\hat{H}_{\mathrm{GR}}$} & \textbf{$\hat{H}_{\mathrm{R}}$} \\ 
\midrule
\hline
(1.185$\pm$0.165) & (1.190$\pm$0.157) & (1.188$\pm$0.178) & (1.167$\pm$0.175) \\ 
\hline
\hline
\bottomrule
\end{tabular}}
\end{centering}
\caption{Mean $R_{\gamma}$ metric ($\pm$ one standard deviation) statistics over $10^3$ gradient-optimised DRAG protocols. \label{tab:RMetric}}
\end{table}

\begin{figure}
    \centering
    % \scalebox{0.95}{\GOATtrajectories}
    \includegraphics[width=0.93\linewidth]{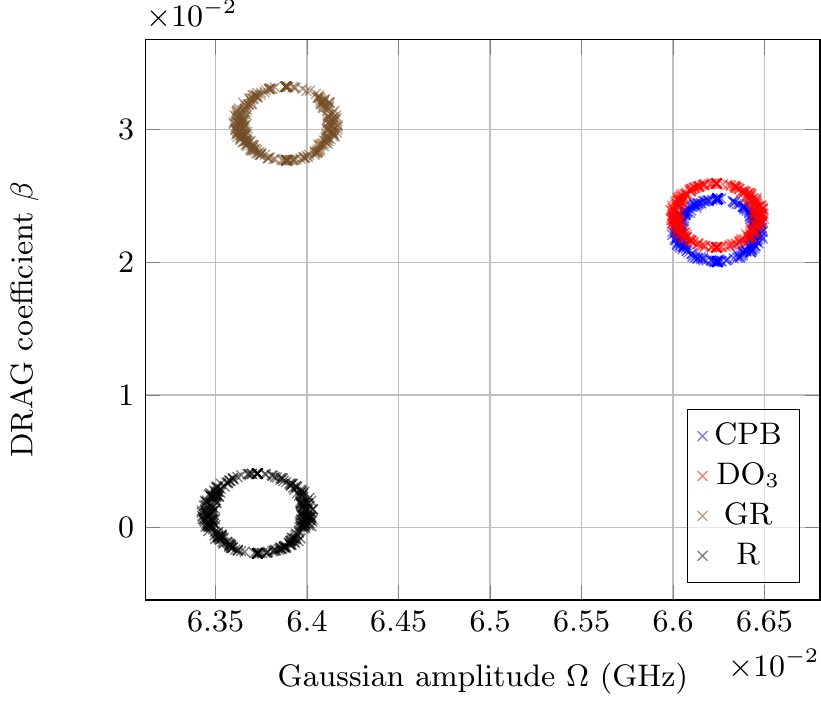}
    \caption{The endpoint distribution of control trajectories generated by GOAT. Initial points in parameter space sampled from a multivariate Gaussian distribution around a previously known maximum are provided to the GOAT algorithm for each model. The blue and red represent the trajectory endpoints of the $\hat{H}_{\mathrm{CPB}}$ and $\hat{H}_{\mathrm{DO_3}}$ models respectively, the distributions of which are near-identical in location and in shape. Models $\hat{H}_{\mathrm{GR}}$ and $\hat{H}_{\mathrm{R}}$, in gold and black, respectively, differ significantly in parameter-space location, but share a similar elliptical distribution. These results demonstrate consistency in the nature of solutions across models, but considerable differences in the parameter-space regions at which they occur.}
    \label{fig:controllability}
\end{figure}

These results indicate that the complexity of landscape navigation is effectively invariant across the models presented, at least under the DRAG pulse parametrisation. This invariance demonstrates that local landscape features are consistent, and can be combined with earlier results to determine that optima location is the key symptom of model deviation. This conclusion can be validated by observing the control trajectory endpoints produced by the gradient-control algorithm. These endpoints are structurally identical for each model, terminating on the surface of an elliptical parameter space that had a large range in DRAG coefficient $\beta$-space and a small range in Gaussian amplitude $\Omega$-space. This behaviour is demonstrated in Fig. \ref{fig:controllability}, omitting trajectories that began with initial points within the termination parameter space.

The resemblance in local landscape structure that is shared across models for standard control techniques invites a future research direction for the development of a model-transfer mapping function. If control landscapes for simple conventional control protocols possess similar local curvature features, it could be reasonably expected that an empirical mapping function could reliably learn to translate a high-population location in parameter space for a simplified model, to the equivalent location of a more accurate model. In light of analysis in Sec. \ref{sec:origins}, within conventional resonant (tightly distributed around the qubit frequency) control settings, this proposed transfer function would approach a diagonal matrix scaling the dimensions of control. Moving beyond resonant control, the mapping function would acquire off-diagonal contributions. In this event a mapping function could, for example, be parametrised by a neural network tasked with learning the nontrivial transfer features across models. This would allow control optimisation algorithms to operate in low-precision simulations and cast their results to the same parameter space in more accurate models, aiming to provide a reduction in computational resource requirements of up to 16 times as discussed in Appendix \ref{appendix:d}. Such a solution could conceivably be most valuable in accelerating a many-qubit control simulation while mitigating crosstalk from a crowded spectrum.\\

\section{Conclusion}

We have presented and characterised four relevant and distinct models of a coupled transmon-resonator architecture. Examining their respective dynamics in a single-qubit control scenario, we have concluded that the reduction in required computational resources of the commonly used generalised quantum Rabi model $\hat{H}_{\mathrm{GR}}$ and two-level quantum Rabi model $\hat{H}_{\mathrm{R}}$ comes at the cost of significant accuracy loss in hardware simulation. We have demonstrated loss in hardware fidelity across models through randomised benchmarking processes on a cloud quantum device for a simple qubit control scheme. We have provided insight into the origins of these deviations, presented simulations of a particularly pathological class of errors, and produced recommendations to optimal control practitioners based on these findings. We have also characterised the models in the context of standard control parameter spaces, presenting both an analysis methodology for and insight into the difficulty of optimal control for each. 

% In general, we note that the selection of Hamiltonian models in previous quantum control research has lacked justification.

% This observation, in conjunction with the results presented in this research, illustrate potential limitations in applying heavily approximated results to hardware.
%In these instances the selection of Hamiltonian models
% of which 
%$E_J/E_C$ operational regime of the transmon qubit is ambiguous
% previous quantum control research has operated in 

Contemporary superconducting quantum devices operate in regimes where the validity of $E_J/E_C\gg 1$ is ambiguous, naturally obscuring the optimal choice of effective Hamiltonian. This investigation provides guidelines for the use of a common set of Hamiltonians, illustrates the loss in fidelity which can be experienced by being liberal with the application of common assumptions, and provides insight into the nature of deviations between these models, all within a set of energy parameters which are experimentally relevant for noisy intermediate-scale quantum devices.

Burgeoning techniques for precise quantum control rely on software which simulates the unitary evolution of a quantum system as accurately as possible, with auxiliary channels responsible for noise and other non-unitary processes \cite{Youssry_2020}. The deployment of learnings from this investigation will maximise the overall performance of these techniques. Further natural extensions of this work arise in more stringently evaluating the validity of these results for the multiqubit case, and investigating the efficacy of the proposed model-transfer mapping function.

\section*{\label{sec:Acknowledgements} Acknowledgements}
We acknowledge the IBM Quantum team for helpful discussions, and access to devices for the experimental portion of this research. We also acknowledge the team behind QuTip, a package used extensively for the simulation portion of this research. We thank the anonymous referees responsible for providing valuable input on this manuscript. Finally, we thank Dr. Maciej Trzaskowski and the wider management at Max Kelsen for support and direction.

\appendix
\renewcommand{\thesection}{\Alph{section}.\arabic{section}}
\renewcommand{\thesubsection}{\thesection\arabic{subsection}}
\setcounter{section}{0}

\begin{appendices}

\section{Nearest-neighbour coupling \label{appendix:a}}
With this approximation, the cavity coupling term can be re-expressed by calculating matrix elements of the number operator in the basis of the uncoupled transmon eigenstates $\ket{i}$,
\begin{equation}
    \hat{n}^{(t)}_{ij} = \bra{j}\hat{n}^{(c)} \ket{i}.
\end{equation}
The superscripts $t$ and $c$ denote the transmon basis and charge basis respectively. In the asymptotic limit of large $E_J/E_C$, perturbation theory for the matrix elements yields,
\begin{align}
    \bra{j+1}\hat{n}\ket{j} &\approx \sqrt{j+1} \left( \frac{1}{2\sqrt{\eta}}\right) \\
    \bra{j+k}\hat{n}\ket{j} &\approx 0,
\end{align}
where $k > 1$. From this, we make the approximation that resonator-induced coupling between transmon states is limited to nearest neighbours. 

\section{Hardware characterisation \label{appendix:b}}

The design of optimal Gaussian pulses for deployment on a real device requires careful characterisation of said device. We utilise Qiskit Pulse access to the \textsf{IBMQ} backend \texttt{ibmq\_armonk} to obtain these parameters. A significant volume of the code used in this process is adapted from the IBM Qiskit textbook \cite{Qiskit-Textbook}.

\subsection{Qubit spectroscopy}
\textsf{IBMQ} backends provide regularly updated estimates of qubit frequencies through their API; nonetheless, the characterisation approach we take includes a qubit spectroscopy process at the beginning of each calibration to ensure the integrity of the subsequent steps. To perform this process, Gaussian pulses are individually mixed with carrier signals at 50 frequencies evenly distributed in a 50 MHz region around the estimated qubit frequency and applied to the qubit. A Lorentzian distribution is fitted to the resultant transmission data to approximate the point of peak transmission (qubit frequency). 

\subsection{$\pi$-pulse calibration}
For further characterisation, a $\pi$-pulse that can be used to reliably flip between computational states is required. For this calibration, a carrier signal is fixed at the previously established qubit frequency, whilst a constant-time (of the order 10$^{-1}$ $\mu$s) Gaussian was swept over a range of pulse amplitudes (on the order of MHz). A sinusoidal fit to the resultant transmission data yielded a Rabi oscillation period, half of which can be adopted as an optimal $\pi$-pulse amplitude.

\subsection{$\ket{e}-\ket{f}$ spectroscopy}
In order to find the second excited level transition frequency, one can apply the same principle as qubit spectroscopy with an initial state of the first excited level $\ket{e}$. At the time of experimentation, the configuration of IBM's Qiskit Pulse package was such that only a single frequency could be associated with a drive line, which prohibited the intuitive protocol of a $\pi$-pulse followed by a sweep of carrier frequencies (mixed with Gaussian pulses) over the frequency region in which the second excited transition is expected. As such, a method of controlling the frequency of the signal indirectly through the drive pulse envelope is required.\\

To do this, the second Gaussian drive pulse (to excite the higher level transition) is further mixed with the sinusoid $e^{2i\pi\Delta ft}$ to generate sidebands at $f_c\pm\Delta f$ (where $f_c$ is the frequency of the carrier signal). Mixing a carrier signal at the qubit frequency with a $\pi$-pulse followed by a $\Delta f$ sweep of sideband pulses allows the second excited transition frequency to be obtained through a Lorentzian fit of transmission data, as in qubit spectroscopy.

\subsection{Dispersive shift}
The dispersive shift of a transmon-resonator system materialises as half the difference between the frequencies of the $\ket{g}$ and $\ket{e}$ transmission peaks. As such, a $\frac{\pi}{2}$-pulse is used to prepare the transmon in a superposition state, and the measurement pulse is swept over a sufficiently fine range of frequencies in order to resolve the two separate transmission peaks apparent at $\omega_{qb} - \chi$ and $\omega_{qb} + \chi$ (corresponding to $\ket{g}$ and $\ket{e}$, respectively). This resolution is achieved for the transmon system by sweeping the resonator frequency over 50 points within a 500 kHz range around the raw resonator frequency.

\subsection{System parameters}
Energy parameters $E_J$, $E_C$, $g$ and $\omega_r$ are required for the modelling approaches detailed in this article. Resonator frequency $\omega_r$ is provided by the Qiskit API. Methods for system parameter estimation vary depending on the model in use. \\

For the Rabi models, the process is simplified by the employment of relations that can be derived from the Rabi set of assumptions. Transmon energy parameters $E_C$ and $E_J$ are estimated from the above measurements using Rabi relations $\omega_{\ket{e}\rightarrow\ket{f}} - \omega_{\ket{g}\rightarrow\ket{e}} = E_C$ and $\omega_{\ket{g}\rightarrow\ket{e}} = \sqrt{8E_CE_J} - E_C$. The coupling strength $g$ is estimated by the dispersive regime relation $\chi = -g^2\frac{E_C}{\Delta(\Delta - E_C)}$.\\

For the CPB and $\mathrm{DO_3}$ models, solutions for $E_J$, $E_C$, and $g$ that return the measured spectrum and dispersive shift are calculated numerically, beginning in the neighbourhood of the parameters estimated by using Rabi relations. 

\section{Perturbation or sextic omission? \label{appendix:b.5}}
Results in Sec. \ref{sub:char_results} highlight deviations in model predictions that manifest in the transition between $\hat{H}_{\mathrm{DO_3}}$ and $\hat{H}_{\mathrm{GR}}$. This divergence presents itself when comparing the dynamics of each model under a Gaussian drive of period 142.2 ns. The procedure to obtain $\hat{H}_{\mathrm{GR}}$ from $\hat{H}_{\mathrm{DO_3}}$ relies on neglecting the higher-order sextic term in the latter model, and subsequently applying a first-order perturbative approximation to the associated quartic term. To verify that the model deviations are largely explainable by the perturbative approximation, the model $\hat{H}_{\mathrm{GR_3}}$ is introduced. This Hamiltonian replicates $\hat{H}_{\mathrm{GR}}$, while additionally including leading-order corrections for the sextic term from $\hat{H}_{\mathrm{DO_3}}$, expressed as
\begin{align}
\label{eq:ham_qr3}
\hat{H}_{\mathrm{GR_3}} &= \sqrt{8E_CE_J}(\hat{b}^{\dagger}\hat{b} + \frac{1}{2}) - E_J \nonumber\\
              &-\hat{E}_{m_3}  + \omega_r \hat{a}^{\dag} \hat{a} + gi(\hat{a} +\hat{a}^\dag)(\hat{b}^\dag -\hat{b}),
\end{align}
where in contrast to Eq. (\ref{eq:em}), the eigenenergy corrections introduced by $\hat{E}_{m_3}$ include third-order terms, 

\begin{align}
\hat{E}_{m_3} &= \frac{E_J}{720}\left(\frac{2E_C}{E_J}\right)^{3/2}(20(\hat{b}^{\dagger}\hat{b})^3 + 30(\hat{b}^{\dagger}\hat{b})^2 + 40\hat{b}^{\dagger}\hat{b} + 15) \nonumber \\
        &-\frac{E_C}{12}(6(\hat{b}^{\dagger}\hat{b})^2 + 6\hat{b}^{\dagger}\hat{b} + 3).
\end{align}

The $\mathrm{DO_3}$, $\mathrm{GR_3}$ and $\mathrm{GR}$ models are each probed by an identical Gaussian pulse mixed with a carrier signal at their respective frequencies, as in Sec. \ref{sub:char_results}. The driving pulse uses a period of 142.2 ns, with amplitudes ranging from 0--75 MHz.

To underscore the model deviations between these intermediate models, we introduce the variable 
\begin{align}
    \Delta_{\mathrm{M_2}}^{\mathrm{M_1}} &= P_{\mathrm{M_1}} - P_{\mathrm{M_2}},
\end{align}
where $P_{\mathrm{M_1}}$ and $P_{\mathrm{M_2}}$ denote the first excited level populations as calculated in the models M$_1$ and M$_2$. The differences $\Delta_{\mathrm{M_2}}^{\mathrm{M_1}}$ quantify the influence of the associated approximations on the system dynamics. 

\begin{figure}[!hbtp]
    \centering
    \includegraphics[width=0.93\linewidth]{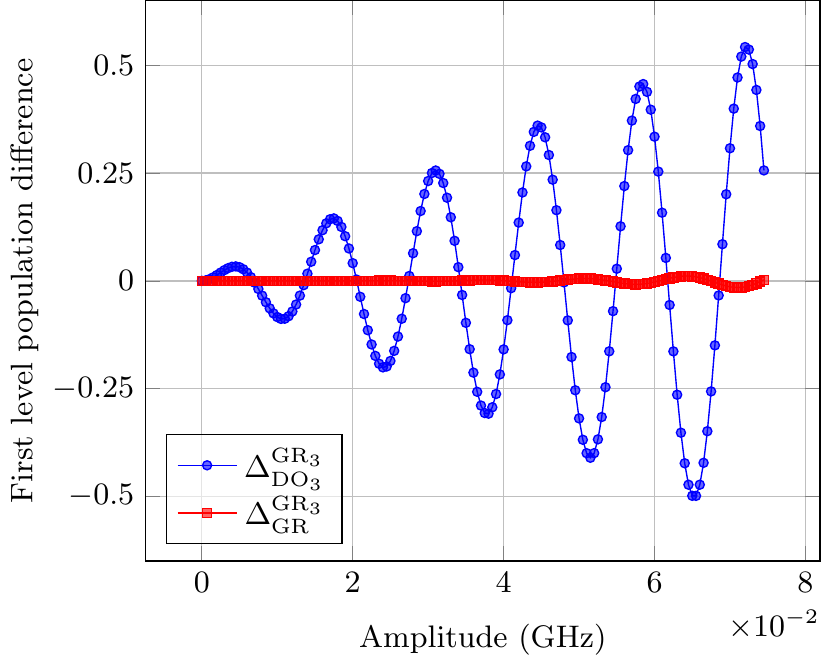}
    % \EjcIDiffInset
    \caption{The behaviour of the metric $\Delta_{\mathrm{M_2}}^{\mathrm{M_1}}$ under application of a 142.2 ns Gaussian driving pulse, plotted with respect to the amplitude of said driving pulse. This metric measures the difference in simulated first energy level population between models M$_1$ and M$_2$. In blue, substantial differences between models $\mathrm{GR_3}$ and $\mathrm{DO_3}$ can be observed, with the difference between these models being the application of a first-order perturbative approximation. In comparison, the red points display the relatively small differences between models $\mathrm{GR_3}$ and $\mathrm{GR}$, attributable to omission of the sextic perturbative term in simulation. This justifies the statement that the first-order perturbative correction in deriving $\hat{H}_\mathrm{GR}$ from $\hat{H}_\mathrm{DO_3}$ is responsible for the magnitude of divergence in simulated dynamics between these models.} 
    \label{fig:gr3}
\end{figure}

The behaviours of $\Delta_{\mathrm{DO_3}}^{\mathrm{GR_3}}$ and $\Delta_{\mathrm{GR}}^{\mathrm{GR_3}}$ across the defined range of driving protocols are demonstrated in Fig. \ref{fig:gr3}. The effects of neglecting the sextic term are trivial in comparison to applying the first-order perturbative approximation, illustrated in red and blue respectively. The validity of the perturbative approximation (predicated on $E_J/E_C \gg 1$) is therefore taken to be the dominant assumption when obtaining $\hat{H}_\mathrm{GR}$ from $\hat{H}_\mathrm{DO_3}$.

\section{Origins of deviations in the frequency domain \label{appendix:c.5}}

\begin{figure}[!hbtp]
    \centering
    \includegraphics[width=0.9\linewidth]{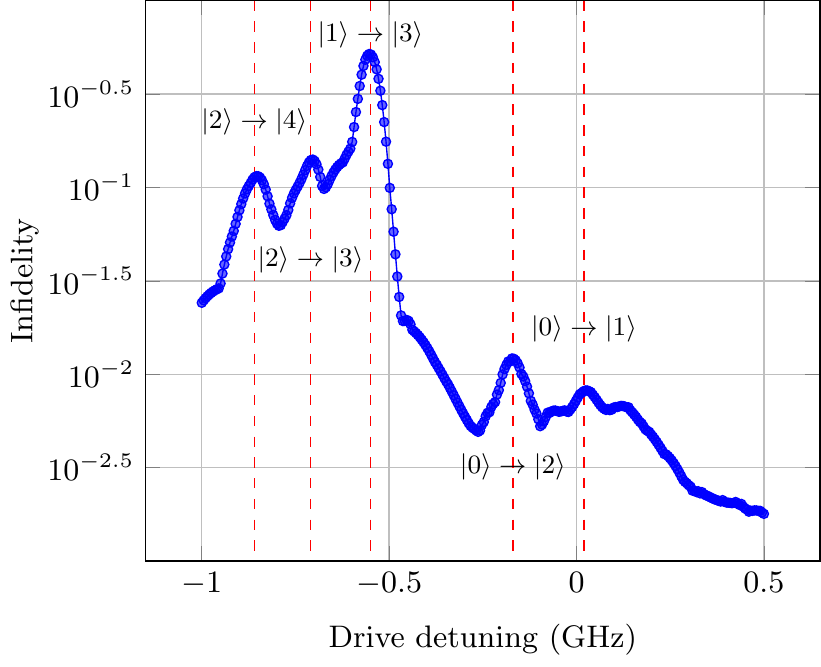}
    % \scalebox{0.9}{\FreqCrossSection}
    \caption{Deviations in simulated control between frequency-matched and amplitude-scaled DO$_3$ and GR models at $E_J/E_C \sim 100$. The $x$ axis is detuning of a 5 ns square drive pulse from the qubit frequency. Amplitudes around 190 MHz are chosen for the drive, scaled to maximally match dynamics. The system begins in a 50-50 superposition between the ground and excited states. The $y$ axis represents the maximum infidelity experienced between models from a subset of times during the evolution of the 5 ns square pulse. Significant infidelity peaks exist at frequencies that correspond to higher-level transitions and two-photon transitions, as well as some small deviation at the qubit frequency. In particular, comparison of the models at the two-photon transition $\ket{1}\rightarrow\ket{3}$ yields an infidelity in excess of 50\%.} 
    \label{fig:fcs}
\end{figure}

The origins of divergences between the DO$_3$ and GR models introduced in Fig. \ref{fig:freq_response} are elucidated in Fig. \ref{fig:fcs}. The parameter calibration in amplitude and frequency leads to a significant mitigation in error at the $\ket{0}\rightarrow\ket{1}$ transition, where maximum infidelity is less than 1\% over the duration of a 5 ns pulse. In comparison, the two-photon transition at $\ket{0}\rightarrow\ket{2}$ increases error to around 1.2\%. Finally, the frequency region that comprises a set of  noncomputational transitions routinely sees infidelities in excess of 10\%, most notably at $\ket{1}\rightarrow\ket{3}$ where a significant portion of the original superposition state is actively driven to the third excited level. This result acts to demonstrate that appropriate amplitude scaling and Stark shift corrections are insufficient measures to ensure a viable control solution, for a control pulse with an arbitrary frequency profile.

Additionally, it is valuable to note that the off-diagonal matrix elements in $\hat{H}_\mathrm{DO_3}$ that are responsible for the dynamical deviations at transitions such as $\ket{0}\rightarrow\ket{2}$ tend towards zero in the limit of $E_J/E_C \rightarrow \infty$. This is a key assumption that allows the first-order perturbation approximation inherent to $\hat{H}_\mathrm{GR}$. The clear discrepancies in these models at $E_J/E_C  \sim 100$, a relatively large experimental ratio, further highlights the questionable justification of this approximation for standard experimental energy regimes.

\section{Numerical simulation \label{appendix:c}}
All model simulations (propagations of time-dependent Hamiltonians) are performed using the zvode ordinary differential equation (ODE) solver provided in Python package Qutip \cite{Johansson_2012}. The solver is deployed with an absolute tolerance of $10^{-8}$, relative tolerance of $10 ^{-6}$, and $5\times10^3$ time steps.\\

The accuracy of numerical simulation of quantum systems can be inhibited by the number of energy levels used. To determine the required Hilbert space to accurately simulate dynamics for each model, we drive an $N$-level transmon from $\ket{0}$ to $\ket{\psi_F}$ with a standard Gaussian pulse of experimentally relevant parameters, from $N = 2$ to $N = 20$ until an error tolerance criteria is met. This criteria is the mean absolute population difference in the computational subspace between $n$ and $n-1$ across the duration of the pulse $T$ (which consists of $f$ timesteps, [$t_0 = 0$, $t_f = T$]). This can be expressed as 

\begin{equation}
    d \left(p_m(n,t) \right) =   \frac{1}{f}\sum_{j=0}^{f} |p_m(n,t=t_j) - p_m(n-1,t=t_j)|,
\end{equation}
where $p_m(n,t)$ returns a vector of the model \textit{m} populations with $n$ energy levels at time step $t$. The model is deemed to converge when this metric reaches 10$^{-5}$. For $\hat{H}_{\mathrm{CPB}}$, $\hat{H}_{\mathrm{DO_3}}$ and $\hat{H}_{\mathrm{GR}}$, required Hilbert space dimensionality was found to be 13, 12, and 6, respectively. Model $\hat{H}_{\mathrm{R}}$ requires only two energy levels by definition.

\section{Computational requirements \label{appendix:d}}

\begin{figure}[!hbtp]
    \centering
    \includegraphics[width=0.9\linewidth]{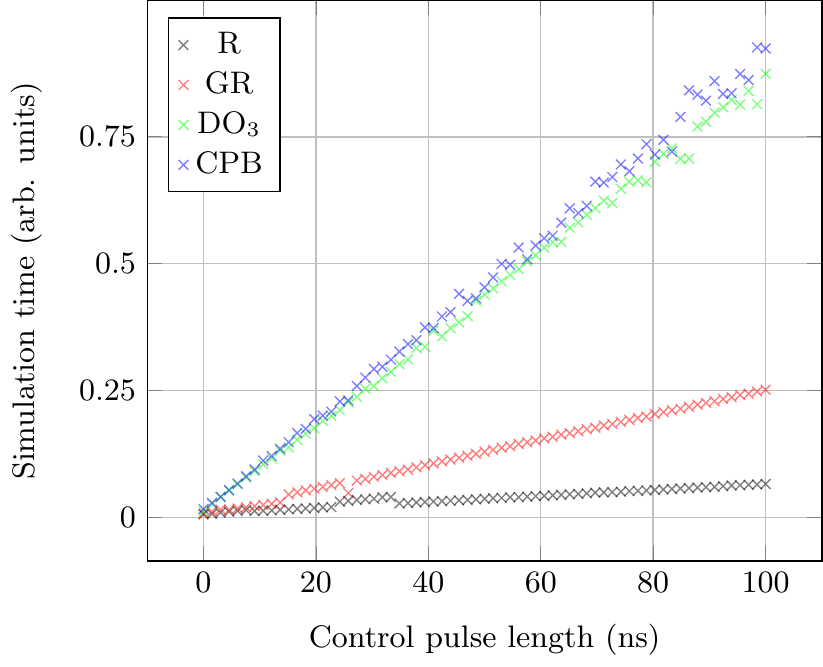}
    % \scalebox{0.9}{\timingplot}
    \caption{Normalised simulation times for Gaussian control pulses of varying periods. The most faithful model (CPB) is seen in blue. The proposed transmon basis model of highest precision ($\mathrm{DO_3}$) is seen in green, while the generalised quantum Rabi (GR) and two-level quantum Rabi (R) models offer significant accelerations in computation time as seen in red and black respectively. This result communicates that the merit of the Rabi family of models (GR and R) lies in their relatively low computational demand.}
    \label{fig:TimingPlot}
\end{figure}

Given the iterative optimisation protocols often used for quantum control, a balance must be struck between the model fidelity and computational footprint. As resource requirements will be largely constrained by the model Hilbert space, the approach detailed in Appendix \ref{appendix:c} is used to define the number of energy levels required for each model. To examine the computational resource requirements of each model, each simulated transmon is driven at its transition frequency by a Gaussian pulse of amplitude 75 MHz, with period ranging between 0 and 150 ns. Linearly fitting these results provides characteristic runtime statistics for each model, which in conjunction with results presented in the body of this paper yields a clear picture of the balance between accuracy and resource requirement. Figure \ref{fig:TimingPlot} illustrates these runtime statistics, using a \texttt{zvode}-centric ODE solver.

Computational efficiency is a clear advantage of the simplified models (GR and R), attributable to both the smaller Hilbert space they require and their diagonal matrix structure. Linear fits reveal that the $\mathrm{DO_3}$ modelling approach offers a marginal 6\% runtime advantage over the full CPB model, in comparison to GR and R offering accelerations of 3.6x and 16.3x respectively. For the set of problems in which precise dynamics is expendable in favour of simulation runtime, Fig. \ref{fig:TimingPlot} is presented in order to demonstrate the value of the simplified quantum Rabi models.\\

\end{appendices}

\bibliography{article}

\begin{thebibliography}{40}%
\makeatletter
\providecommand \@ifxundefined [1]{%
 \@ifx{#1\undefined}
}%
\providecommand \@ifnum [1]{%
 \ifnum #1\expandafter \@firstoftwo
 \else \expandafter \@secondoftwo
 \fi
}%
\providecommand \@ifx [1]{%
 \ifx #1\expandafter \@firstoftwo
 \else \expandafter \@secondoftwo
 \fi
}%
\providecommand \natexlab [1]{#1}%
\providecommand \enquote  [1]{``#1''}%
\providecommand \bibnamefont  [1]{#1}%
\providecommand \bibfnamefont [1]{#1}%
\providecommand \citenamefont [1]{#1}%
\providecommand \href@noop [0]{\@secondoftwo}%
\providecommand \href [0]{\begingroup \@sanitize@url \@href}%
\providecommand \@href[1]{\@@startlink{#1}\@@href}%
\providecommand \@@href[1]{\endgroup#1\@@endlink}%
\providecommand \@sanitize@url [0]{\catcode `\\12\catcode `\$12\catcode
  `\&12\catcode `\#12\catcode `\^12\catcode `\_12\catcode `\%12\relax}%
\providecommand \@@startlink[1]{}%
\providecommand \@@endlink[0]{}%
\providecommand \url  [0]{\begingroup\@sanitize@url \@url }%
\providecommand \@url [1]{\endgroup\@href {#1}{\urlprefix }}%
\providecommand \urlprefix  [0]{URL }%
\providecommand \Eprint [0]{\href }%
\providecommand \doibase [0]{http://dx.doi.org/}%
\providecommand \selectlanguage [0]{\@gobble}%
\providecommand \bibinfo  [0]{\@secondoftwo}%
\providecommand \bibfield  [0]{\@secondoftwo}%
\providecommand \translation [1]{[#1]}%
\providecommand \BibitemOpen [0]{}%
\providecommand \bibitemStop [0]{}%
\providecommand \bibitemNoStop [0]{.\EOS\space}%
\providecommand \EOS [0]{\spacefactor3000\relax}%
\providecommand \BibitemShut  [1]{\csname bibitem#1\endcsname}%
\let\auto@bib@innerbib\@empty
%</preamble>
\bibitem [{\citenamefont {Allen}\ \emph {et~al.}(2017)\citenamefont {Allen},
  \citenamefont {Kosut}, \citenamefont {Joo}, \citenamefont {Leek},\ and\
  \citenamefont {Ginossar}}]{Allen_2017}%
  \BibitemOpen
  \bibfield  {author} {\bibinfo {author} {\bibfnamefont {Joseph~L.}\
  \bibnamefont {Allen}}, \bibinfo {author} {\bibfnamefont {Robert}\
  \bibnamefont {Kosut}}, \bibinfo {author} {\bibfnamefont {Jaewoo}\
  \bibnamefont {Joo}}, \bibinfo {author} {\bibfnamefont {Peter}\ \bibnamefont
  {Leek}}, \ and\ \bibinfo {author} {\bibfnamefont {Eran}\ \bibnamefont
  {Ginossar}},\ }\bibfield  {title} {\enquote {\bibinfo {title} {Optimal
  control of two qubits via a single cavity drive in circuit quantum
  electrodynamics},}\ }\href {\doibase 10.1103/PhysRevA.95.042325} {\bibfield
  {journal} {\bibinfo  {journal} {Phys. Rev. A}\ }\textbf {\bibinfo {volume}
  {95}},\ \bibinfo {pages} {042325} (\bibinfo {year} {2017})}\BibitemShut
  {NoStop}%
\bibitem [{\citenamefont {Spiteri}\ \emph {et~al.}(2018)\citenamefont
  {Spiteri}, \citenamefont {Schmidt}, \citenamefont {Ghosh}, \citenamefont
  {Zahedinejad},\ and\ \citenamefont {Sanders}}]{Spiteri_2018}%
  \BibitemOpen
  \bibfield  {author} {\bibinfo {author} {\bibfnamefont {Raymond~J}\
  \bibnamefont {Spiteri}}, \bibinfo {author} {\bibfnamefont {Marina}\
  \bibnamefont {Schmidt}}, \bibinfo {author} {\bibfnamefont {Joydip}\
  \bibnamefont {Ghosh}}, \bibinfo {author} {\bibfnamefont {Ehsan}\ \bibnamefont
  {Zahedinejad}}, \ and\ \bibinfo {author} {\bibfnamefont {Barry~C}\
  \bibnamefont {Sanders}},\ }\bibfield  {title} {\enquote {\bibinfo {title}
  {Quantum control for high-fidelity multi-qubit gates},}\ }\href {\doibase
  10.1088/1367-2630/aae79a} {\bibfield  {journal} {\bibinfo  {journal} {New J.
  Phys.}\ }\textbf {\bibinfo {volume} {20}},\ \bibinfo {pages} {113009}
  (\bibinfo {year} {2018})}\BibitemShut {NoStop}%
\bibitem [{\citenamefont {Werninghaus}\ \emph {et~al.}(2021)\citenamefont
  {Werninghaus}, \citenamefont {Egger}, \citenamefont {Roy}, \citenamefont
  {Machnes}, \citenamefont {Wilhelm},\ and\ \citenamefont
  {Filipp}}]{Werninghaus_2021}%
  \BibitemOpen
  \bibfield  {author} {\bibinfo {author} {\bibfnamefont {M.}~\bibnamefont
  {Werninghaus}}, \bibinfo {author} {\bibfnamefont {D.~J.}\ \bibnamefont
  {Egger}}, \bibinfo {author} {\bibfnamefont {F.}~\bibnamefont {Roy}}, \bibinfo
  {author} {\bibfnamefont {S.}~\bibnamefont {Machnes}}, \bibinfo {author}
  {\bibfnamefont {F.~K.}\ \bibnamefont {Wilhelm}}, \ and\ \bibinfo {author}
  {\bibfnamefont {S.}~\bibnamefont {Filipp}},\ }\bibfield  {title} {\enquote
  {\bibinfo {title} {Leakage reduction in fast superconducting qubit gates via
  optimal control},}\ }\href {https://doi.org/10.1038/s41534-020-00346-2}
  {\bibfield  {journal} {\bibinfo  {journal} {NPJ Quantum Inf.}\ }\textbf
  {\bibinfo {volume} {7}},\ \bibinfo {pages} {14} (\bibinfo {year}
  {2021})}\BibitemShut {NoStop}%
\bibitem [{\citenamefont {Kirchhoff}\ \emph {et~al.}(2018)\citenamefont
  {Kirchhoff}, \citenamefont {Ke\ss{}ler}, \citenamefont {Liebermann},
  \citenamefont {Ass\'emat}, \citenamefont {Machnes}, \citenamefont {Motzoi},\
  and\ \citenamefont {Wilhelm}}]{Kirchhoff_2018}%
  \BibitemOpen
  \bibfield  {author} {\bibinfo {author} {\bibfnamefont {Susanna}\ \bibnamefont
  {Kirchhoff}}, \bibinfo {author} {\bibfnamefont {Torsten}\ \bibnamefont
  {Ke\ss{}ler}}, \bibinfo {author} {\bibfnamefont {Per~J.}\ \bibnamefont
  {Liebermann}}, \bibinfo {author} {\bibfnamefont {Elie}\ \bibnamefont
  {Ass\'emat}}, \bibinfo {author} {\bibfnamefont {Shai}\ \bibnamefont
  {Machnes}}, \bibinfo {author} {\bibfnamefont {Felix}\ \bibnamefont {Motzoi}},
  \ and\ \bibinfo {author} {\bibfnamefont {Frank~K.}\ \bibnamefont {Wilhelm}},\
  }\bibfield  {title} {\enquote {\bibinfo {title} {Optimized cross-resonance
  gate for coupled transmon systems},}\ }\href
  {https://link.aps.org/doi/10.1103/PhysRevA.97.042348} {\bibfield  {journal}
  {\bibinfo  {journal} {Phys. Rev. A}\ }\textbf {\bibinfo {volume} {97}},\
  \bibinfo {pages} {042348} (\bibinfo {year} {2018})}\BibitemShut {NoStop}%
\bibitem [{\citenamefont {Wu}\ \emph {et~al.}(2020)\citenamefont {Wu},
  \citenamefont {Tomarken}, \citenamefont {Petersson}, \citenamefont
  {Martinez}, \citenamefont {Rosen},\ and\ \citenamefont {DuBois}}]{Wu_2020}%
  \BibitemOpen
  \bibfield  {author} {\bibinfo {author} {\bibfnamefont {Xian}\ \bibnamefont
  {Wu}}, \bibinfo {author} {\bibfnamefont {S.~L.}\ \bibnamefont {Tomarken}},
  \bibinfo {author} {\bibfnamefont {N.~Anders}\ \bibnamefont {Petersson}},
  \bibinfo {author} {\bibfnamefont {L.~A.}\ \bibnamefont {Martinez}}, \bibinfo
  {author} {\bibfnamefont {Yaniv~J.}\ \bibnamefont {Rosen}}, \ and\ \bibinfo
  {author} {\bibfnamefont {Jonathan~L.}\ \bibnamefont {DuBois}},\ }\bibfield
  {title} {\enquote {\bibinfo {title} {High-fidelity software-defined quantum
  logic on a superconducting qudit},}\ }\href
  {https://link.aps.org/doi/10.1103/PhysRevLett.125.170502} {\bibfield
  {journal} {\bibinfo  {journal} {Phys. Rev. Lett.}\ }\textbf {\bibinfo
  {volume} {125}},\ \bibinfo {pages} {170502} (\bibinfo {year}
  {2020})}\BibitemShut {NoStop}%
\bibitem [{\citenamefont {Garc\'{\i}a-Ripoll}\ \emph
  {et~al.}(2020)\citenamefont {Garc\'{\i}a-Ripoll}, \citenamefont
  {Ruiz-Chamorro},\ and\ \citenamefont {Torrontegui}}]{Garcia_2020}%
  \BibitemOpen
  \bibfield  {author} {\bibinfo {author} {\bibfnamefont {J.~J.}\ \bibnamefont
  {Garc\'{\i}a-Ripoll}}, \bibinfo {author} {\bibfnamefont {A.}~\bibnamefont
  {Ruiz-Chamorro}}, \ and\ \bibinfo {author} {\bibfnamefont {E.}~\bibnamefont
  {Torrontegui}},\ }\bibfield  {title} {\enquote {\bibinfo {title} {Quantum
  control of frequency-tunable transmon superconducting qubits},}\ }\href
  {\doibase 10.1103/PhysRevApplied.14.044035} {\bibfield  {journal} {\bibinfo
  {journal} {Phys. Rev. Appl.}\ }\textbf {\bibinfo {volume} {14}},\ \bibinfo
  {pages} {044035} (\bibinfo {year} {2020})}\BibitemShut {NoStop}%
\bibitem [{\citenamefont {Riaz}\ \emph {et~al.}(2019)\citenamefont {Riaz},
  \citenamefont {Shuang},\ and\ \citenamefont
  {Qamar}}]{riazOptimalControlMethods2019}%
  \BibitemOpen
  \bibfield  {author} {\bibinfo {author} {\bibfnamefont {Bilal}\ \bibnamefont
  {Riaz}}, \bibinfo {author} {\bibfnamefont {Cong}\ \bibnamefont {Shuang}}, \
  and\ \bibinfo {author} {\bibfnamefont {Shahid}\ \bibnamefont {Qamar}},\
  }\bibfield  {title} {\enquote {\bibinfo {title} {Optimal control methods for
  quantum gate preparation: a comparative study},}\ }\href
  {https://doi.org/10.1007/s11128-019-2190-0} {\bibfield  {journal} {\bibinfo
  {journal} {Quantum Inf. Process.}\ }\textbf {\bibinfo {volume} {18}},\
  \bibinfo {pages} {100} (\bibinfo {year} {2019})}\BibitemShut {NoStop}%
\bibitem [{\citenamefont {An}\ and\ \citenamefont {Zhou}(2019)}]{An_2019}%
  \BibitemOpen
  \bibfield  {author} {\bibinfo {author} {\bibfnamefont {Zheng}\ \bibnamefont
  {An}}\ and\ \bibinfo {author} {\bibfnamefont {D.~L.}\ \bibnamefont {Zhou}},\
  }\bibfield  {title} {\enquote {\bibinfo {title} {Deep reinforcement learning
  for quantum gate control},}\ }\href
  {https://doi.org/10.1209/0295-5075/126/60002} {\bibfield  {journal} {\bibinfo
   {journal} {EPL}\ }\textbf {\bibinfo {volume} {126}},\ \bibinfo {pages}
  {60002} (\bibinfo {year} {2019})}\BibitemShut {NoStop}%
\bibitem [{\citenamefont {Zhang}\ \emph {et~al.}(2019)\citenamefont {Zhang},
  \citenamefont {Wei}, \citenamefont {Asad}, \citenamefont {Yang},\ and\
  \citenamefont {Wang}}]{Zhang2019}%
  \BibitemOpen
  \bibfield  {author} {\bibinfo {author} {\bibfnamefont {Xiao-Ming}\
  \bibnamefont {Zhang}}, \bibinfo {author} {\bibfnamefont {Zezhu}\ \bibnamefont
  {Wei}}, \bibinfo {author} {\bibfnamefont {Raza}\ \bibnamefont {Asad}},
  \bibinfo {author} {\bibfnamefont {Xu-Chen}\ \bibnamefont {Yang}}, \ and\
  \bibinfo {author} {\bibfnamefont {Xin}\ \bibnamefont {Wang}},\ }\bibfield
  {title} {\enquote {\bibinfo {title} {When does reinforcement learning stand
  out in quantum control? a comparative study on state preparation},}\ }\href
  {https://doi.org/10.1038/s41534-019-0201-8} {\bibfield  {journal} {\bibinfo
  {journal} {NPJ Quantum Inf.}\ }\textbf {\bibinfo {volume} {5}},\ \bibinfo
  {pages} {85} (\bibinfo {year} {2019})}\BibitemShut {NoStop}%
\bibitem [{\citenamefont {Xu}\ \emph {et~al.}(2019)\citenamefont {Xu},
  \citenamefont {Li}, \citenamefont {Liu}, \citenamefont {Wang}, \citenamefont
  {Yuan},\ and\ \citenamefont {Wang}}]{Xu_2019}%
  \BibitemOpen
  \bibfield  {author} {\bibinfo {author} {\bibfnamefont {Han}\ \bibnamefont
  {Xu}}, \bibinfo {author} {\bibfnamefont {Junning}\ \bibnamefont {Li}},
  \bibinfo {author} {\bibfnamefont {Liqiang}\ \bibnamefont {Liu}}, \bibinfo
  {author} {\bibfnamefont {Yu}~\bibnamefont {Wang}}, \bibinfo {author}
  {\bibfnamefont {Haidong}\ \bibnamefont {Yuan}}, \ and\ \bibinfo {author}
  {\bibfnamefont {Xin}\ \bibnamefont {Wang}},\ }\bibfield  {title} {\enquote
  {\bibinfo {title} {Generalizable control for quantum parameter estimation
  through reinforcement learning},}\ }\href
  {https://doi.org/10.1038/s41534-019-0198-z} {\bibfield  {journal} {\bibinfo
  {journal} {NPJ Quantum Inf.}\ }\textbf {\bibinfo {volume} {5}},\ \bibinfo
  {pages} {82} (\bibinfo {year} {2019})}\BibitemShut {NoStop}%
\bibitem [{\citenamefont {Wauters}\ \emph {et~al.}(2020)\citenamefont
  {Wauters}, \citenamefont {Panizon}, \citenamefont {Mbeng},\ and\
  \citenamefont {Santoro}}]{Wauters_2020}%
  \BibitemOpen
  \bibfield  {author} {\bibinfo {author} {\bibfnamefont {Matteo~M.}\
  \bibnamefont {Wauters}}, \bibinfo {author} {\bibfnamefont {Emanuele}\
  \bibnamefont {Panizon}}, \bibinfo {author} {\bibfnamefont {Glen~B.}\
  \bibnamefont {Mbeng}}, \ and\ \bibinfo {author} {\bibfnamefont {Giuseppe~E.}\
  \bibnamefont {Santoro}},\ }\bibfield  {title} {\enquote {\bibinfo {title}
  {Reinforcement-learning-assisted quantum optimization},}\ }\href
  {http://dx.doi.org/10.1103/PhysRevResearch.2.033446} {\bibfield  {journal}
  {\bibinfo  {journal} {Phys. Rev. Research}\ }\textbf {\bibinfo {volume}
  {2}},\ \bibinfo {pages} {033446} (\bibinfo {year} {2020})}\BibitemShut
  {NoStop}%
\bibitem [{\citenamefont {Zahedinejad}\ \emph {et~al.}(2016)\citenamefont
  {Zahedinejad}, \citenamefont {Ghosh},\ and\ \citenamefont
  {Sanders}}]{Zahedinejad_2016}%
  \BibitemOpen
  \bibfield  {author} {\bibinfo {author} {\bibfnamefont {Ehsan}\ \bibnamefont
  {Zahedinejad}}, \bibinfo {author} {\bibfnamefont {Joydip}\ \bibnamefont
  {Ghosh}}, \ and\ \bibinfo {author} {\bibfnamefont {Barry~C.}\ \bibnamefont
  {Sanders}},\ }\bibfield  {title} {\enquote {\bibinfo {title} {Designing
  high-fidelity single-shot three-qubit gates: A machine-learning approach},}\
  }\href {http://dx.doi.org/10.1103/PhysRevApplied.6.054005} {\bibfield
  {journal} {\bibinfo  {journal} {Phys. Rev. Appl.}\ }\textbf {\bibinfo
  {volume} {6}},\ \bibinfo {pages} {054005} (\bibinfo {year}
  {2016})}\BibitemShut {NoStop}%
\bibitem [{\citenamefont {Niu}\ \emph {et~al.}(2019)\citenamefont {Niu},
  \citenamefont {Boixo}, \citenamefont {Smelyanskiy},\ and\ \citenamefont
  {Neven}}]{Niu_2019}%
  \BibitemOpen
  \bibfield  {author} {\bibinfo {author} {\bibfnamefont {Murphy~Yeuzhen}\
  \bibnamefont {Niu}}, \bibinfo {author} {\bibfnamefont {Sergio}\ \bibnamefont
  {Boixo}}, \bibinfo {author} {\bibfnamefont {Vadim~N.}\ \bibnamefont
  {Smelyanskiy}}, \ and\ \bibinfo {author} {\bibfnamefont {Hartmut}\
  \bibnamefont {Neven}},\ }\bibfield  {title} {\enquote {\bibinfo {title}
  {Universal quantum control through deep reinforcement learning},}\ }\href
  {https://doi.org/10.1038/s41534-019-0141-3} {\bibfield  {journal} {\bibinfo
  {journal} {NPJ Quantum Inf.}\ }\textbf {\bibinfo {volume} {5}},\ \bibinfo
  {pages} {33} (\bibinfo {year} {2019})}\BibitemShut {NoStop}%
\bibitem [{\citenamefont {Bukov}\ \emph {et~al.}(2018)\citenamefont {Bukov},
  \citenamefont {Day}, \citenamefont {Sels}, \citenamefont {Weinberg},
  \citenamefont {Polkovnikov},\ and\ \citenamefont {Mehta}}]{Bukov_2018}%
  \BibitemOpen
  \bibfield  {author} {\bibinfo {author} {\bibfnamefont {Marin}\ \bibnamefont
  {Bukov}}, \bibinfo {author} {\bibfnamefont {Alexandre G.~R.}\ \bibnamefont
  {Day}}, \bibinfo {author} {\bibfnamefont {Dries}\ \bibnamefont {Sels}},
  \bibinfo {author} {\bibfnamefont {Phillip}\ \bibnamefont {Weinberg}},
  \bibinfo {author} {\bibfnamefont {Anatoli}\ \bibnamefont {Polkovnikov}}, \
  and\ \bibinfo {author} {\bibfnamefont {Pankaj}\ \bibnamefont {Mehta}},\
  }\bibfield  {title} {\enquote {\bibinfo {title} {Reinforcement learning in
  different phases of quantum control},}\ }\href
  {http://dx.doi.org/10.1103/PhysRevX.8.031086} {\bibfield  {journal} {\bibinfo
   {journal} {Phys. Rev. X}\ }\textbf {\bibinfo {volume} {8}},\ \bibinfo
  {pages} {031086} (\bibinfo {year} {2018})}\BibitemShut {NoStop}%
\bibitem [{\citenamefont {Wu}\ \emph {et~al.}(2019)\citenamefont {Wu},
  \citenamefont {Ding}, \citenamefont {Dong},\ and\ \citenamefont
  {Wang}}]{Wu_2019}%
  \BibitemOpen
  \bibfield  {author} {\bibinfo {author} {\bibfnamefont {Re-Bing}\ \bibnamefont
  {Wu}}, \bibinfo {author} {\bibfnamefont {Haijin}\ \bibnamefont {Ding}},
  \bibinfo {author} {\bibfnamefont {Daoyi}\ \bibnamefont {Dong}}, \ and\
  \bibinfo {author} {\bibfnamefont {Xiaoting}\ \bibnamefont {Wang}},\
  }\bibfield  {title} {\enquote {\bibinfo {title} {Learning robust and
  high-precision quantum controls},}\ }\href
  {http://dx.doi.org/10.1103/PhysRevA.99.042327} {\bibfield  {journal}
  {\bibinfo  {journal} {Phys. Rev. A}\ }\textbf {\bibinfo {volume} {99}},\
  \bibinfo {pages} {042327} (\bibinfo {year} {2019})}\BibitemShut {NoStop}%
\bibitem [{\citenamefont {Abdelhafez}\ \emph {et~al.}(2020)\citenamefont
  {Abdelhafez}, \citenamefont {Baker}, \citenamefont {Gyenis}, \citenamefont
  {Mundada}, \citenamefont {Houck}, \citenamefont {Schuster},\ and\
  \citenamefont {Koch}}]{Abdelhafez_2020}%
  \BibitemOpen
  \bibfield  {author} {\bibinfo {author} {\bibfnamefont {Mohamed}\ \bibnamefont
  {Abdelhafez}}, \bibinfo {author} {\bibfnamefont {Brian}\ \bibnamefont
  {Baker}}, \bibinfo {author} {\bibfnamefont {Andr\'as}\ \bibnamefont
  {Gyenis}}, \bibinfo {author} {\bibfnamefont {Pranav}\ \bibnamefont
  {Mundada}}, \bibinfo {author} {\bibfnamefont {Andrew~A.}\ \bibnamefont
  {Houck}}, \bibinfo {author} {\bibfnamefont {David}\ \bibnamefont {Schuster}},
  \ and\ \bibinfo {author} {\bibfnamefont {Jens}\ \bibnamefont {Koch}},\
  }\bibfield  {title} {\enquote {\bibinfo {title} {Universal gates for
  protected superconducting qubits using optimal control},}\ }\href {\doibase
  10.1103/PhysRevA.101.022321} {\bibfield  {journal} {\bibinfo  {journal}
  {Phys. Rev. A}\ }\textbf {\bibinfo {volume} {101}},\ \bibinfo {pages}
  {022321} (\bibinfo {year} {2020})}\BibitemShut {NoStop}%
\bibitem [{\citenamefont {Koch}\ \emph {et~al.}(2007)\citenamefont {Koch},
  \citenamefont {Yu}, \citenamefont {Gambetta}, \citenamefont {Houck},
  \citenamefont {Schuster}, \citenamefont {Majer}, \citenamefont {Blais},
  \citenamefont {Devoret}, \citenamefont {Girvin},\ and\ \citenamefont
  {Schoelkopf}}]{Koch_2007}%
  \BibitemOpen
  \bibfield  {author} {\bibinfo {author} {\bibfnamefont {Jens}\ \bibnamefont
  {Koch}}, \bibinfo {author} {\bibfnamefont {Terri~M.}\ \bibnamefont {Yu}},
  \bibinfo {author} {\bibfnamefont {Jay}\ \bibnamefont {Gambetta}}, \bibinfo
  {author} {\bibfnamefont {A.~A.}\ \bibnamefont {Houck}}, \bibinfo {author}
  {\bibfnamefont {D.~I.}\ \bibnamefont {Schuster}}, \bibinfo {author}
  {\bibfnamefont {J.}~\bibnamefont {Majer}}, \bibinfo {author} {\bibfnamefont
  {Alexandre}\ \bibnamefont {Blais}}, \bibinfo {author} {\bibfnamefont {M.~H.}\
  \bibnamefont {Devoret}}, \bibinfo {author} {\bibfnamefont {S.~M.}\
  \bibnamefont {Girvin}}, \ and\ \bibinfo {author} {\bibfnamefont {R.~J.}\
  \bibnamefont {Schoelkopf}},\ }\bibfield  {title} {\enquote {\bibinfo {title}
  {Charge-insensitive qubit design derived from the cooper pair box},}\ }\href
  {http://dx.doi.org/10.1103/PhysRevA.76.042319} {\bibfield  {journal}
  {\bibinfo  {journal} {Phys. Rev. A}\ }\textbf {\bibinfo {volume} {76}},\
  \bibinfo {pages} {042319} (\bibinfo {year} {2007})}\BibitemShut {NoStop}%
\bibitem [{\citenamefont {Motzoi}\ \emph {et~al.}(2009)\citenamefont {Motzoi},
  \citenamefont {Gambetta}, \citenamefont {Rebentrost},\ and\ \citenamefont
  {Wilhelm}}]{motzoi_gambetta_rebentrost_wilhelm_2009}%
  \BibitemOpen
  \bibfield  {author} {\bibinfo {author} {\bibfnamefont {F.}~\bibnamefont
  {Motzoi}}, \bibinfo {author} {\bibfnamefont {J.~M.}\ \bibnamefont
  {Gambetta}}, \bibinfo {author} {\bibfnamefont {P.}~\bibnamefont
  {Rebentrost}}, \ and\ \bibinfo {author} {\bibfnamefont {F.~K.}\ \bibnamefont
  {Wilhelm}},\ }\bibfield  {title} {\enquote {\bibinfo {title} {Simple pulses
  for elimination of leakage in weakly nonlinear qubits},}\ }\href
  {https://journals.aps.org/prl/abstract/10.1103/PhysRevLett.103.110501}
  {\bibfield  {journal} {\bibinfo  {journal} {Phys. Rev. Lett.}\ }\textbf
  {\bibinfo {volume} {103}},\ \bibinfo {pages} {110501} (\bibinfo {year}
  {2009})}\BibitemShut {NoStop}%
\bibitem [{\citenamefont {Reich}\ \emph {et~al.}(2012)\citenamefont {Reich},
  \citenamefont {Ndong},\ and\ \citenamefont {Koch}}]{Reich_2012}%
  \BibitemOpen
  \bibfield  {author} {\bibinfo {author} {\bibfnamefont {Daniel~M.}\
  \bibnamefont {Reich}}, \bibinfo {author} {\bibfnamefont {Mamadou}\
  \bibnamefont {Ndong}}, \ and\ \bibinfo {author} {\bibfnamefont
  {Christiane~P.}\ \bibnamefont {Koch}},\ }\bibfield  {title} {\enquote
  {\bibinfo {title} {Monotonically convergent optimization in quantum control
  using krotov’s method},}\ }\href {http://dx.doi.org/10.1063/1.3691827}
  {\bibfield  {journal} {\bibinfo  {journal} {J. Chem. Phys.}\ }\textbf
  {\bibinfo {volume} {136}},\ \bibinfo {pages} {104103} (\bibinfo {year}
  {2012})}\BibitemShut {NoStop}%
\bibitem [{\citenamefont {Machnes}\ \emph {et~al.}(2018)\citenamefont
  {Machnes}, \citenamefont {Assémat}, \citenamefont {Tannor},\ and\
  \citenamefont {Wilhelm}}]{Machnes_2018}%
  \BibitemOpen
  \bibfield  {author} {\bibinfo {author} {\bibfnamefont {Shai}\ \bibnamefont
  {Machnes}}, \bibinfo {author} {\bibfnamefont {Elie}\ \bibnamefont
  {Assémat}}, \bibinfo {author} {\bibfnamefont {David}\ \bibnamefont
  {Tannor}}, \ and\ \bibinfo {author} {\bibfnamefont {Frank~K.}\ \bibnamefont
  {Wilhelm}},\ }\bibfield  {title} {\enquote {\bibinfo {title} {Tunable,
  flexible, and efficient optimization of control pulses for practical
  qubits},}\ }\href {http://dx.doi.org/10.1103/PhysRevLett.120.150401}
  {\bibfield  {journal} {\bibinfo  {journal} {Phys. Rev. Lett.}\ }\textbf
  {\bibinfo {volume} {120}},\ \bibinfo {pages} {150401} (\bibinfo {year}
  {2018})}\BibitemShut {NoStop}%
\bibitem [{\citenamefont {Khaneja}\ \emph {et~al.}(2004)\citenamefont
  {Khaneja}, \citenamefont {Reiss}, \citenamefont {Kehlet}, \citenamefont
  {Schulte-Herbrüggen},\ and\ \citenamefont {Glaser}}]{Khaneja_2004}%
  \BibitemOpen
  \bibfield  {author} {\bibinfo {author} {\bibfnamefont {Navin}\ \bibnamefont
  {Khaneja}}, \bibinfo {author} {\bibfnamefont {Timo}\ \bibnamefont {Reiss}},
  \bibinfo {author} {\bibfnamefont {Cindie}\ \bibnamefont {Kehlet}}, \bibinfo
  {author} {\bibfnamefont {Thomas}\ \bibnamefont {Schulte-Herbrüggen}}, \ and\
  \bibinfo {author} {\bibfnamefont {Steffen~J.}\ \bibnamefont {Glaser}},\
  }\bibfield  {title} {\enquote {\bibinfo {title} {Optimal control of coupled
  spin dynamics: design of nmr pulse sequences by gradient ascent
  algorithms},}\ }\href {https://doi.org/10.1016/j.jmr.2004.11.004} {\bibfield
  {journal} {\bibinfo  {journal} {J. Magn. Reson.}\ }\textbf {\bibinfo {volume}
  {172}},\ \bibinfo {pages} {296--305} (\bibinfo {year} {2004})}\BibitemShut
  {NoStop}%
\bibitem [{\citenamefont {Liebermann}\ and\ \citenamefont
  {Wilhelm}(2016)}]{Liebermann_2016}%
  \BibitemOpen
  \bibfield  {author} {\bibinfo {author} {\bibfnamefont {Per~J.}\ \bibnamefont
  {Liebermann}}\ and\ \bibinfo {author} {\bibfnamefont {Frank~K.}\ \bibnamefont
  {Wilhelm}},\ }\bibfield  {title} {\enquote {\bibinfo {title} {Optimal qubit
  control using single-flux quantum pulses},}\ }\href
  {http://dx.doi.org/10.1103/PhysRevApplied.6.024022} {\bibfield  {journal}
  {\bibinfo  {journal} {Phys. Rev. Appl.}\ }\textbf {\bibinfo {volume} {6}},\
  \bibinfo {pages} {024022} (\bibinfo {year} {2016})}\BibitemShut {NoStop}%
\bibitem [{\citenamefont {Daraeizadeh}\ \emph {et~al.}(2020)\citenamefont
  {Daraeizadeh}, \citenamefont {Premaratne}, \citenamefont {Khammassi},
  \citenamefont {Song}, \citenamefont {Perkowski},\ and\ \citenamefont
  {Matsuura}}]{Daraeizadeh_2020}%
  \BibitemOpen
  \bibfield  {author} {\bibinfo {author} {\bibfnamefont {S.}~\bibnamefont
  {Daraeizadeh}}, \bibinfo {author} {\bibfnamefont {S.~P.}\ \bibnamefont
  {Premaratne}}, \bibinfo {author} {\bibfnamefont {N.}~\bibnamefont
  {Khammassi}}, \bibinfo {author} {\bibfnamefont {X.}~\bibnamefont {Song}},
  \bibinfo {author} {\bibfnamefont {M.}~\bibnamefont {Perkowski}}, \ and\
  \bibinfo {author} {\bibfnamefont {A.~Y.}\ \bibnamefont {Matsuura}},\
  }\bibfield  {title} {\enquote {\bibinfo {title} {Machine-learning-based
  three-qubit gate design for the toffoli gate and parity check in transmon
  systems},}\ }\href {http://dx.doi.org/10.1103/PhysRevA.102.012601} {\bibfield
   {journal} {\bibinfo  {journal} {Phys. Rev. A}\ }\textbf {\bibinfo {volume}
  {102}},\ \bibinfo {pages} {012601} (\bibinfo {year} {2020})}\BibitemShut
  {NoStop}%
\bibitem [{\citenamefont {Basilewitsch}\ \emph {et~al.}(2019)\citenamefont
  {Basilewitsch}, \citenamefont {Cosco}, \citenamefont {Lo~Gullo},
  \citenamefont {Möttönen}, \citenamefont {Ala-Nissilä}, \citenamefont
  {Koch},\ and\ \citenamefont {Maniscalco}}]{Basilewitsch_2019}%
  \BibitemOpen
  \bibfield  {author} {\bibinfo {author} {\bibfnamefont {Daniel}\ \bibnamefont
  {Basilewitsch}}, \bibinfo {author} {\bibfnamefont {Francesco}\ \bibnamefont
  {Cosco}}, \bibinfo {author} {\bibfnamefont {Nicolino}\ \bibnamefont
  {Lo~Gullo}}, \bibinfo {author} {\bibfnamefont {Mikko}\ \bibnamefont
  {Möttönen}}, \bibinfo {author} {\bibfnamefont {Tapio}\ \bibnamefont
  {Ala-Nissilä}}, \bibinfo {author} {\bibfnamefont {Christiane~P}\
  \bibnamefont {Koch}}, \ and\ \bibinfo {author} {\bibfnamefont {Sabrina}\
  \bibnamefont {Maniscalco}},\ }\bibfield  {title} {\enquote {\bibinfo {title}
  {Reservoir engineering using quantum optimal control for qubit reset},}\
  }\href {http://dx.doi.org/10.1088/1367-2630/ab41ad} {\bibfield  {journal}
  {\bibinfo  {journal} {New J. Phys.}\ }\textbf {\bibinfo {volume} {21}},\
  \bibinfo {pages} {093054} (\bibinfo {year} {2019})}\BibitemShut {NoStop}%
\bibitem [{\citenamefont {Ballester}\ \emph {et~al.}(2012)\citenamefont
  {Ballester}, \citenamefont {Romero}, \citenamefont {Garc\'{\i}a-Ripoll},
  \citenamefont {Deppe},\ and\ \citenamefont {Solano}}]{Ballester_2012}%
  \BibitemOpen
  \bibfield  {author} {\bibinfo {author} {\bibfnamefont {D.}~\bibnamefont
  {Ballester}}, \bibinfo {author} {\bibfnamefont {G.}~\bibnamefont {Romero}},
  \bibinfo {author} {\bibfnamefont {J.~J.}\ \bibnamefont {Garc\'{\i}a-Ripoll}},
  \bibinfo {author} {\bibfnamefont {F.}~\bibnamefont {Deppe}}, \ and\ \bibinfo
  {author} {\bibfnamefont {E.}~\bibnamefont {Solano}},\ }\bibfield  {title}
  {\enquote {\bibinfo {title} {Quantum simulation of the ultrastrong-coupling
  dynamics in circuit quantum electrodynamics},}\ }\href {\doibase
  10.1103/PhysRevX.2.021007} {\bibfield  {journal} {\bibinfo  {journal} {Phys.
  Rev. X}\ }\textbf {\bibinfo {volume} {2}},\ \bibinfo {pages} {021007}
  (\bibinfo {year} {2012})}\BibitemShut {NoStop}%
\bibitem [{\citenamefont {Bao}\ \emph {et~al.}(2018)\citenamefont {Bao},
  \citenamefont {Kleer}, \citenamefont {Wang},\ and\ \citenamefont
  {Rahmani}}]{Bao_2018}%
  \BibitemOpen
  \bibfield  {author} {\bibinfo {author} {\bibfnamefont {Seraph}\ \bibnamefont
  {Bao}}, \bibinfo {author} {\bibfnamefont {Silken}\ \bibnamefont {Kleer}},
  \bibinfo {author} {\bibfnamefont {Ruoyu}\ \bibnamefont {Wang}}, \ and\
  \bibinfo {author} {\bibfnamefont {Armin}\ \bibnamefont {Rahmani}},\
  }\bibfield  {title} {\enquote {\bibinfo {title} {Optimal control of
  superconducting gmon qubits using pontryagin{\textquotesingle}s minimum
  principle: Preparing a maximally entangled state with singular bang-bang
  protocols},}\ }\href {https://doi.org/10.1103/physreva.97.062343} {\bibfield
  {journal} {\bibinfo  {journal} {Phys. Rev. A}\ }\textbf {\bibinfo {volume}
  {97}},\ \bibinfo {pages} {062343} (\bibinfo {year} {2018})}\BibitemShut
  {NoStop}%
\bibitem [{\citenamefont {Willsch}\ \emph {et~al.}(2017)\citenamefont
  {Willsch}, \citenamefont {Nocon}, \citenamefont {Jin}, \citenamefont
  {Raedt},\ and\ \citenamefont {Michielsen}}]{Willsch2017}%
  \BibitemOpen
  \bibfield  {author} {\bibinfo {author} {\bibfnamefont {D.}~\bibnamefont
  {Willsch}}, \bibinfo {author} {\bibfnamefont {M.}~\bibnamefont {Nocon}},
  \bibinfo {author} {\bibfnamefont {F.}~\bibnamefont {Jin}}, \bibinfo {author}
  {\bibfnamefont {H.~De}\ \bibnamefont {Raedt}}, \ and\ \bibinfo {author}
  {\bibfnamefont {K.}~\bibnamefont {Michielsen}},\ }\bibfield  {title}
  {\enquote {\bibinfo {title} {Gate-error analysis in simulations of quantum
  computers with transmon qubits},}\ }\href {\doibase
  10.1103/physreva.96.062302} {\bibfield  {journal} {\bibinfo  {journal}
  {Physical Review A}\ }\textbf {\bibinfo {volume} {96}} (\bibinfo {year}
  {2017}),\ 10.1103/physreva.96.062302}\BibitemShut {NoStop}%
\bibitem [{\citenamefont {Magesan}\ \emph {et~al.}(2011)\citenamefont
  {Magesan}, \citenamefont {Gambetta},\ and\ \citenamefont
  {Emerson}}]{Magesan_2011}%
  \BibitemOpen
  \bibfield  {author} {\bibinfo {author} {\bibfnamefont {Easwar}\ \bibnamefont
  {Magesan}}, \bibinfo {author} {\bibfnamefont {J.~M.}\ \bibnamefont
  {Gambetta}}, \ and\ \bibinfo {author} {\bibfnamefont {Joseph}\ \bibnamefont
  {Emerson}},\ }\bibfield  {title} {\enquote {\bibinfo {title} {Scalable and
  robust randomized benchmarking of quantum processes},}\ }\href
  {http://dx.doi.org/10.1103/PhysRevLett.106.180504} {\bibfield  {journal}
  {\bibinfo  {journal} {Phys. Rev. Lett.}\ }\textbf {\bibinfo {volume} {106}},\
  \bibinfo {pages} {180504} (\bibinfo {year} {2011})}\BibitemShut {NoStop}%
\bibitem [{\citenamefont {Nanduri}\ \emph {et~al.}(2013)\citenamefont
  {Nanduri}, \citenamefont {Donovan}, \citenamefont {Ho},\ and\ \citenamefont
  {Rabitz}}]{Nanduri_2013}%
  \BibitemOpen
  \bibfield  {author} {\bibinfo {author} {\bibfnamefont {Arun}\ \bibnamefont
  {Nanduri}}, \bibinfo {author} {\bibfnamefont {Ashley}\ \bibnamefont
  {Donovan}}, \bibinfo {author} {\bibfnamefont {Tak-San}\ \bibnamefont {Ho}}, \
  and\ \bibinfo {author} {\bibfnamefont {Herschel}\ \bibnamefont {Rabitz}},\
  }\bibfield  {title} {\enquote {\bibinfo {title} {Exploring quantum control
  landscape structure},}\ }\href {http://dx.doi.org/10.1103/PhysRevA.88.033425}
  {\bibfield  {journal} {\bibinfo  {journal} {Phys. Rev. A}\ }\textbf {\bibinfo
  {volume} {88}},\ \bibinfo {pages} {033425} (\bibinfo {year}
  {2013})}\BibitemShut {NoStop}%
\bibitem [{\citenamefont {Chow}\ \emph {et~al.}(2013)\citenamefont {Chow},
  \citenamefont {Gambetta}, \citenamefont {Cross}, \citenamefont {Merkel},
  \citenamefont {Rigetti},\ and\ \citenamefont {Steffen}}]{Chow2013}%
  \BibitemOpen
  \bibfield  {author} {\bibinfo {author} {\bibfnamefont {Jerry~M}\ \bibnamefont
  {Chow}}, \bibinfo {author} {\bibfnamefont {Jay~M}\ \bibnamefont {Gambetta}},
  \bibinfo {author} {\bibfnamefont {Andrew~W}\ \bibnamefont {Cross}}, \bibinfo
  {author} {\bibfnamefont {Seth~T}\ \bibnamefont {Merkel}}, \bibinfo {author}
  {\bibfnamefont {Chad}\ \bibnamefont {Rigetti}}, \ and\ \bibinfo {author}
  {\bibfnamefont {M}~\bibnamefont {Steffen}},\ }\bibfield  {title} {\enquote
  {\bibinfo {title} {Microwave-activated conditional-phase gate for
  superconducting qubits},}\ }\href
  {https://doi.org/10.1088/1367-2630/15/11/115012} {\bibfield  {journal}
  {\bibinfo  {journal} {New Journal of Physics}\ }\textbf {\bibinfo {volume}
  {15}},\ \bibinfo {pages} {115012} (\bibinfo {year} {2013})}\BibitemShut
  {NoStop}%
\bibitem [{\citenamefont {Magesan}\ and\ \citenamefont
  {Gambetta}(2020)}]{Magesan2020}%
  \BibitemOpen
  \bibfield  {author} {\bibinfo {author} {\bibfnamefont {Easwar}\ \bibnamefont
  {Magesan}}\ and\ \bibinfo {author} {\bibfnamefont {Jay~M.}\ \bibnamefont
  {Gambetta}},\ }\bibfield  {title} {\enquote {\bibinfo {title} {Effective
  hamiltonian models of the cross-resonance gate},}\ }\href
  {https://doi.org/10.1103/physreva.101.052308} {\bibfield  {journal} {\bibinfo
   {journal} {Physical Review A}\ }\textbf {\bibinfo {volume} {101}} (\bibinfo
  {year} {2020})}\BibitemShut {NoStop}%
\bibitem [{\citenamefont {Neg{\^{\i}}rneac}\ \emph {et~al.}(2021)\citenamefont
  {Neg{\^{\i}}rneac}, \citenamefont {Ali}, \citenamefont {Muthusubramanian},
  \citenamefont {Battistel}, \citenamefont {Sagastizabal}, \citenamefont
  {Moreira}, \citenamefont {Marques}, \citenamefont {Vlothuizen}, \citenamefont
  {Beekman}, \citenamefont {Zachariadis}, \citenamefont {Haider}, \citenamefont
  {Bruno},\ and\ \citenamefont {DiCarlo}}]{Negrneac2021}%
  \BibitemOpen
  \bibfield  {author} {\bibinfo {author} {\bibfnamefont {V.}~\bibnamefont
  {Neg{\^{\i}}rneac}}, \bibinfo {author} {\bibfnamefont {H.}~\bibnamefont
  {Ali}}, \bibinfo {author} {\bibfnamefont {N.}~\bibnamefont
  {Muthusubramanian}}, \bibinfo {author} {\bibfnamefont {F.}~\bibnamefont
  {Battistel}}, \bibinfo {author} {\bibfnamefont {R.}~\bibnamefont
  {Sagastizabal}}, \bibinfo {author} {\bibfnamefont {M.{\hspace{0.167em}}S.}\
  \bibnamefont {Moreira}}, \bibinfo {author} {\bibfnamefont
  {J.{\hspace{0.167em}}F.}\ \bibnamefont {Marques}}, \bibinfo {author}
  {\bibfnamefont {W.{\hspace{0.167em}}J.}\ \bibnamefont {Vlothuizen}}, \bibinfo
  {author} {\bibfnamefont {M.}~\bibnamefont {Beekman}}, \bibinfo {author}
  {\bibfnamefont {C.}~\bibnamefont {Zachariadis}}, \bibinfo {author}
  {\bibfnamefont {N.}~\bibnamefont {Haider}}, \bibinfo {author} {\bibfnamefont
  {A.}~\bibnamefont {Bruno}}, \ and\ \bibinfo {author} {\bibfnamefont
  {L.}~\bibnamefont {DiCarlo}},\ }\bibfield  {title} {\enquote {\bibinfo
  {title} {High-fidelity controlled- z gate with maximal intermediate leakage
  operating at the speed limit in a superconducting quantum processor},}\
  }\href {https://doi.org/10.1103/physrevlett.126.220502} {\bibfield  {journal}
  {\bibinfo  {journal} {Physical Review Letters}\ }\textbf {\bibinfo {volume}
  {126}} (\bibinfo {year} {2021})}\BibitemShut {NoStop}%
\bibitem [{\citenamefont {Chakrabarti}\ and\ \citenamefont
  {Rabitz}(2007)}]{Chakrabarti2007}%
  \BibitemOpen
  \bibfield  {author} {\bibinfo {author} {\bibfnamefont {Raj}\ \bibnamefont
  {Chakrabarti}}\ and\ \bibinfo {author} {\bibfnamefont {Herschel}\
  \bibnamefont {Rabitz}},\ }\bibfield  {title} {\enquote {\bibinfo {title}
  {Quantum control landscapes},}\ }\href {\doibase 10.1080/01442350701633300}
  {\bibfield  {journal} {\bibinfo  {journal} {Int. Rev. Phys. Chem.}\ }\textbf
  {\bibinfo {volume} {26}},\ \bibinfo {pages} {671--735} (\bibinfo {year}
  {2007})}\BibitemShut {NoStop}%
\bibitem [{\citenamefont {Moore}\ \emph {et~al.}(2008)\citenamefont {Moore},
  \citenamefont {Hsieh},\ and\ \citenamefont {Rabitz}}]{Moore_2008}%
  \BibitemOpen
  \bibfield  {author} {\bibinfo {author} {\bibfnamefont {Katharine}\
  \bibnamefont {Moore}}, \bibinfo {author} {\bibfnamefont {Michael}\
  \bibnamefont {Hsieh}}, \ and\ \bibinfo {author} {\bibfnamefont {Herschel}\
  \bibnamefont {Rabitz}},\ }\bibfield  {title} {\enquote {\bibinfo {title} {On
  the relationship between quantum control landscape structure and optimization
  complexity},}\ }\href {https://doi.org/10.1063/1.2907740} {\bibfield
  {journal} {\bibinfo  {journal} {J. Chem. Phys.}\ }\textbf {\bibinfo {volume}
  {128}},\ \bibinfo {pages} {154117} (\bibinfo {year} {2008})}\BibitemShut
  {NoStop}%
\bibitem [{\citenamefont {Larocca}\ \emph {et~al.}(2018)\citenamefont
  {Larocca}, \citenamefont {Poggi},\ and\ \citenamefont
  {Wisniacki}}]{Larocca_2018}%
  \BibitemOpen
  \bibfield  {author} {\bibinfo {author} {\bibfnamefont {Martín}\ \bibnamefont
  {Larocca}}, \bibinfo {author} {\bibfnamefont {Pablo~M.}\ \bibnamefont
  {Poggi}}, \ and\ \bibinfo {author} {\bibfnamefont {Diego~A.}\ \bibnamefont
  {Wisniacki}},\ }\bibfield  {title} {\enquote {\bibinfo {title} {Quantum
  control landscape for a two-level system near the quantum speed limit},}\
  }\href {https://doi.org/10.1088/1751-8121/aad657} {\bibfield  {journal}
  {\bibinfo  {journal} {J. Phys. A Math. Theor.}\ }\textbf {\bibinfo {volume}
  {51}},\ \bibinfo {pages} {385305} (\bibinfo {year} {2018})}\BibitemShut
  {NoStop}%
\bibitem [{\citenamefont {Day}\ \emph {et~al.}(2019)\citenamefont {Day},
  \citenamefont {Bukov}, \citenamefont {Weinberg}, \citenamefont {Mehta},\ and\
  \citenamefont {Sels}}]{Day_2019}%
  \BibitemOpen
  \bibfield  {author} {\bibinfo {author} {\bibfnamefont {Alexandre G.~R.}\
  \bibnamefont {Day}}, \bibinfo {author} {\bibfnamefont {Marin}\ \bibnamefont
  {Bukov}}, \bibinfo {author} {\bibfnamefont {Phillip}\ \bibnamefont
  {Weinberg}}, \bibinfo {author} {\bibfnamefont {Pankaj}\ \bibnamefont
  {Mehta}}, \ and\ \bibinfo {author} {\bibfnamefont {Dries}\ \bibnamefont
  {Sels}},\ }\bibfield  {title} {\enquote {\bibinfo {title} {Glassy phase of
  optimal quantum control},}\ }\href {\doibase 10.1103/PhysRevLett.122.020601}
  {\bibfield  {journal} {\bibinfo  {journal} {Phys. Rev. Lett.}\ }\textbf
  {\bibinfo {volume} {122}},\ \bibinfo {pages} {020601} (\bibinfo {year}
  {2019})}\BibitemShut {NoStop}%
\bibitem [{\citenamefont {Kosut}\ \emph {et~al.}(2019)\citenamefont {Kosut},
  \citenamefont {Arenz},\ and\ \citenamefont {Rabitz}}]{Kosut2019}%
  \BibitemOpen
  \bibfield  {author} {\bibinfo {author} {\bibfnamefont {Robert~L}\
  \bibnamefont {Kosut}}, \bibinfo {author} {\bibfnamefont {Christian}\
  \bibnamefont {Arenz}}, \ and\ \bibinfo {author} {\bibfnamefont {Herschel}\
  \bibnamefont {Rabitz}},\ }\bibfield  {title} {\enquote {\bibinfo {title}
  {Quantum control landscape of bipartite systems},}\ }\href {\doibase
  10.1088/1751-8121/ab0dc9} {\bibfield  {journal} {\bibinfo  {journal} {J.
  Phys. A Math. Theor.}\ }\textbf {\bibinfo {volume} {52}},\ \bibinfo {pages}
  {165305} (\bibinfo {year} {2019})}\BibitemShut {NoStop}%
\bibitem [{\citenamefont {Youssry}\ \emph {et~al.}(2020)\citenamefont
  {Youssry}, \citenamefont {Paz-Silva},\ and\ \citenamefont
  {Ferrie}}]{Youssry_2020}%
  \BibitemOpen
  \bibfield  {author} {\bibinfo {author} {\bibfnamefont {Akram}\ \bibnamefont
  {Youssry}}, \bibinfo {author} {\bibfnamefont {Gerardo~A.}\ \bibnamefont
  {Paz-Silva}}, \ and\ \bibinfo {author} {\bibfnamefont {Christopher}\
  \bibnamefont {Ferrie}},\ }\bibfield  {title} {\enquote {\bibinfo {title}
  {Characterization and control of open quantum systems beyond quantum noise
  spectroscopy},}\ }\href {http://dx.doi.org/10.1038/s41534-020-00332-8}
  {\bibfield  {journal} {\bibinfo  {journal} {NPJ Quantum Inf.}\ }\textbf
  {\bibinfo {volume} {6}},\ \bibinfo {pages} {95} (\bibinfo {year}
  {2020})}\BibitemShut {NoStop}%
\bibitem [{\citenamefont {Asfaw}\ \emph {et~al.}(2020)\citenamefont {Asfaw},
  \citenamefont {Bello}, \citenamefont {Ben-Haim}, \citenamefont {Bravyi},
  \citenamefont {Bronn}, \citenamefont {Capelluto}, \citenamefont {Vazquez},
  \citenamefont {Ceroni}, \citenamefont {Chen}, \citenamefont {Frisch},
  \citenamefont {Gambetta}, \citenamefont {Garion}, \citenamefont {Gil},
  \citenamefont {Gonzalez}, \citenamefont {Harkins}, \citenamefont {Imamichi},
  \citenamefont {McKay}, \citenamefont {Mezzacapo}, \citenamefont {Minev},
  \citenamefont {Movassagh}, \citenamefont {Nannicni}, \citenamefont {Nation},
  \citenamefont {Phan}, \citenamefont {Pistoia}, \citenamefont {Rattew},
  \citenamefont {Schaefer}, \citenamefont {Shabani}, \citenamefont {Smolin},
  \citenamefont {Temme}, \citenamefont {Tod}, \citenamefont {Wood},\ and\
  \citenamefont {Wootton}}]{Qiskit-Textbook}%
  \BibitemOpen
  \bibfield  {author} {\bibinfo {author} {\bibfnamefont {A.}~\bibnamefont
  {Asfaw}}, \bibinfo {author} {\bibfnamefont {L.}~\bibnamefont {Bello}},
  \bibinfo {author} {\bibfnamefont {Y.}~\bibnamefont {Ben-Haim}}, \bibinfo
  {author} {\bibfnamefont {S.}~\bibnamefont {Bravyi}}, \bibinfo {author}
  {\bibfnamefont {N.}~\bibnamefont {Bronn}}, \bibinfo {author} {\bibfnamefont
  {L.}~\bibnamefont {Capelluto}}, \bibinfo {author} {\bibfnamefont
  {A.~Carrera}\ \bibnamefont {Vazquez}}, \bibinfo {author} {\bibfnamefont
  {J.}~\bibnamefont {Ceroni}}, \bibinfo {author} {\bibfnamefont
  {R.}~\bibnamefont {Chen}}, \bibinfo {author} {\bibfnamefont {A.}~\bibnamefont
  {Frisch}}, \bibinfo {author} {\bibfnamefont {J.}~\bibnamefont {Gambetta}},
  \bibinfo {author} {\bibfnamefont {S.}~\bibnamefont {Garion}}, \bibinfo
  {author} {\bibfnamefont {L.}~\bibnamefont {Gil}}, \bibinfo {author}
  {\bibfnamefont {S.~De La~Puente}\ \bibnamefont {Gonzalez}}, \bibinfo {author}
  {\bibfnamefont {F.}~\bibnamefont {Harkins}}, \bibinfo {author} {\bibfnamefont
  {T.}~\bibnamefont {Imamichi}}, \bibinfo {author} {\bibfnamefont
  {D.}~\bibnamefont {McKay}}, \bibinfo {author} {\bibfnamefont
  {A.}~\bibnamefont {Mezzacapo}}, \bibinfo {author} {\bibfnamefont
  {Z.}~\bibnamefont {Minev}}, \bibinfo {author} {\bibfnamefont
  {R.}~\bibnamefont {Movassagh}}, \bibinfo {author} {\bibfnamefont
  {G.}~\bibnamefont {Nannicni}}, \bibinfo {author} {\bibfnamefont
  {P.}~\bibnamefont {Nation}}, \bibinfo {author} {\bibfnamefont
  {A.}~\bibnamefont {Phan}}, \bibinfo {author} {\bibfnamefont {M.}~\bibnamefont
  {Pistoia}}, \bibinfo {author} {\bibfnamefont {A.}~\bibnamefont {Rattew}},
  \bibinfo {author} {\bibfnamefont {J.}~\bibnamefont {Schaefer}}, \bibinfo
  {author} {\bibfnamefont {J.}~\bibnamefont {Shabani}}, \bibinfo {author}
  {\bibfnamefont {J.}~\bibnamefont {Smolin}}, \bibinfo {author} {\bibfnamefont
  {K.}~\bibnamefont {Temme}}, \bibinfo {author} {\bibfnamefont
  {M.}~\bibnamefont {Tod}}, \bibinfo {author} {\bibfnamefont {S.}~\bibnamefont
  {Wood}}, \ and\ \bibinfo {author} {\bibfnamefont {J.}~\bibnamefont
  {Wootton}},\ }\href {http://community.qiskit.org/textbook} {\enquote
  {\bibinfo {title} {Learn quantum computation using qiskit},}\ } (\bibinfo
  {year} {2020})\BibitemShut {NoStop}%
\bibitem [{\citenamefont {Johansson}\ \emph {et~al.}(2012)\citenamefont
  {Johansson}, \citenamefont {Nation},\ and\ \citenamefont
  {Nori}}]{Johansson_2012}%
  \BibitemOpen
  \bibfield  {author} {\bibinfo {author} {\bibfnamefont {J.R.}\ \bibnamefont
  {Johansson}}, \bibinfo {author} {\bibfnamefont {P.D.}\ \bibnamefont
  {Nation}}, \ and\ \bibinfo {author} {\bibfnamefont {Franco}\ \bibnamefont
  {Nori}},\ }\bibfield  {title} {\enquote {\bibinfo {title} {Qutip: An
  open-source python framework for the dynamics of open quantum systems},}\
  }\href {\doibase 10.1016/j.cpc.2012.02.021} {\bibfield  {journal} {\bibinfo
  {journal} {Comput. Phys. Commun.}\ }\textbf {\bibinfo {volume} {183}},\
  \bibinfo {pages} {1760–1772} (\bibinfo {year} {2012})}\BibitemShut
  {NoStop}%
\end{thebibliography}%

\end{document}